\documentclass[journal]{IEEEtran}
\usepackage{subfigure}
\usepackage{graphicx}
\usepackage{epsfig,latexsym,amsmath,graphics}
\usepackage{amssymb,dsfont,amsthm}
\usepackage{cite,color}
\usepackage{url}
\usepackage{booktabs}
\usepackage{algorithm,algorithmicx}
\usepackage{algpseudocode}
\usepackage{stfloats}
\usepackage{verbatim}
\usepackage{bm}
\usepackage{stfloats}
\usepackage{multirow}
\usepackage{ulem}
\usepackage{multirow}

\begin{document}
	%
	\title{Low-Complexity Rydberg Array Reuse: Modeling and Receiver Design for Sparse Channels}
	%
	%
	%
		%
	
	\author{Hao Wu, Shanchi Wu, Xinyuan Yao, Rui Ni and Chen Gong
		\thanks{This work was supported by National Natural Science Foundation of China	under Grant 62331024 and 62171428.}
		\thanks{Hao Wu, Shanchi Wu, Xinyuan Yao, Chen Gong are with the School of Information Science and Technology in University of Science and Technology of China, Email address: \{wuhao0719, wsc0807, yxy200127\}@mail.ustc.edu.cn, cgong821@ustc.edu.cn.
			
		Rui Ni is with Huawei Technology, Email address: raney.nirui@huawei.com.

}
	}

\maketitle

\begin{abstract}
Rydberg atomic quantum receivers have been seen as novel radio frequency measurements and the high sensitivity to a large range of frequencies makes it attractive for communications reception. However, current implementations of Rydberg array antennas predominantly rely on simple stacking of multiple single-antenna units. While conceptually straightforward, this approach leads to substantial system bulkiness due to the unique requirements of atomic sensors, particularly the need for multiple spatially separated laser setups, rendering such designs both impractical for real-world applications and challenging to fabricate. This limitation underscores the critical need for developing multiplexed Rydberg sensor array architectures. In the domain of conventional RF array antennas, hybrid analog-digital beamforming has emerged as a pivotal architecture for large-scale millimeter-wave (mmWave) multiple-input multiple-output (MIMO) systems, as it substantially reduces the hardware complexity associated with fully-digital beamforming while closely approaching its performance. Drawing inspiration from this methodology, we conduct a systematic study in this work on the design principles, equivalent modeling, and precoding strategies for low-complexity multiplexed Rydberg array, an endeavor crucial to enabling practical and scalable quantum-enhanced communication systems.

\end{abstract}

\section{Introduction}
Rydberg atoms have recently emerged as a promising platform for electric field sensing, enabling direct SI-traceable and self-calibrated measurements \cite{schlossberger2024rydberg,holloway2017electric}. With electrons in highly excited states characterized by large principal quantum numbers, Rydberg atoms exhibit exceptional sensitivity to radio frequency (RF) signal, making them ideally suited for developing atomic-scale sensors for communication signal detection. These capabilities have led to widespread applications across multiple domains, including polarization measurement \cite{sedlacek2013atom}, angle-of-arrival estimation \cite{robinson2021determining,mao2023digital}, subwavelength imaging \cite{downes2020full}, near-field antenna pattern characterization \cite{shi2023new}, and multi-frequency signal recognition \cite{zhang2024image}.

In works \cite{gong2025rydberg, schlossberger2024rydberg}, the authors summarize eight fundamental receiver architectures for single-receiver systems, and propose both single-input single-output (SISO) and multiple-input multiple-output (MIMO) models based on Rydberg sensors. Furthermore, regardless of the reception scheme, the final detection process ultimately relies on photoelectric conversion. Considering the requirements for complex optical paths and avalanche photodiode (APD) in practical large-scale antenna deployment, together with the directional nature of multipath propagation that introduces correlation among antenna array signals, it is essential to analyze the performance of simplified Rydberg arrays, particularly those based on reuse architectures.

\subsection{Related Works in Rydberg Array}
Currently, a number of recent studies have centered on the development of MIMO Rydberg receivers\cite{gong2025rydberg,zhu2025raq,atapattu2025detection,song2025csi,cui2025mimo,xu2025channel,cui2024multi,yuan2025electromagnetic,xu2024design,cui2025towards,kim2025multi,yan2025measurement,wu2024enhancing}. However, existing work predominantly relies on classical RF antenna array models, failing to account for the fundamentally distinct properties of Rydberg arrays. For instance, in large-scale Rydberg array deployments, the overall antenna dimensions are critically constrained by the physical size of atomic vapor cells and the integration density of local oscillator antennas. Therefore, a mature Rydberg MIMO architecture and design must account for both the complexity and miniaturization challenges inherent in practical large-scale array deployment.

The earliest Rydberg antenna units were based on free-space lasers, but their large physical dimensions limited array miniaturization\cite{holloway2014sub,holloway2017electric,sedlacek2013atom}. Works \cite{mao2022high,deb2018radio,simons2018fiber} demonstrated that employing fiber-based devices could significantly reduce the system footprint, thereby enabling large-scale array integration. In Works \cite{jing2020atomic,robinson2021determining}, local oscillator (LO) antennas were shown to effectively enhance detection sensitivity and enable phase information acquisition. However, the relatively large size of these LO antennas posed constraints for miniaturized, large-scale array applications. Works \cite{xie2025low,holloway2022rydberg} addressed this challenge by introducing a parallel electrode plane-based LO design, which effectively reduced the antenna size and facilitated its compact integration with an atomic vapor cell. This configuration established what is now the most prevalent and fundamental architecture: one dedicated LO antenna per atomic vapor cell.

Furthermore, the laser beam diameter (approximately 1 mm) is typically much smaller than the physical size of the atomic vapor cell (1-5 cm)\cite{otto2021data}. Allocating only one laser pair per cell thus leads to significant underutilization of the available volume. Consequently, subsequent integration designs have aimed to incorporate multiple laser pairs within a single atomic vapor cell, thereby increasing the number of antenna elements within a given area\cite{yan2025measurement,wu2024enhancing}. Additionally, to reduce the number of costly APD at the receiver, a design featuring a convex lens that combines the signals from multiple laser pairs within a single cell before detection by a shared APD has been effectively implemented, minimizing the required APD count\cite{wu2024enhancing,otto2021data}.

In summary, reviewing the evolution of Rydberg array design reveals three primary directions for achieving higher integration density and miniaturization: LO signal reuse, atomic vapor cell reuse, and APD reuse.

\subsection{Related Works in MIMO Precoding}

Inspired by the advantages of hybrid precoding, substantial research efforts have been devoted in recent years to its architecture design and optimization algorithms. For fully-connected (FC) and partially-connected (PC) structures, a variety of optimization methods have been developed, including those based on orthogonal matching pursuit (OMP) \cite{el2014spatially}, manifold optimization \cite{lee2014hybrid}, alternating minimization \cite{yu2016alternating}, and singular value decomposition (SVD) \cite{li2017hybrid,zhang2018svd}.

Regarding hybrid precoding with finite-resolution phase shifters (PSs), prior studies\cite{he2023energy,zhu2023max,nie2023spectrum,sohrabi2016hybrid,chen2017hybrid,lyu2021lattice,sohrabi2017hybrid} have explored this area. One common strategy applies direct quantization to the analog precoding matrix derived under infinite-resolution assumptions. Nevertheless, this often introduces substantial quantization errors, particularly detrimental to OMP and SVD-based methods whose performance relies on preserved orthogonality. In response, iterative optimization frameworks have been developed to progressively approach optimal spectral efficiency by adjusting PS phases \cite{sohrabi2016hybrid,sohrabi2017hybrid}. Furthermore, to streamline this process, studies in \cite{chen2017hybrid,lyu2021lattice} transform the optimization objective from complex spectral efficiency to minimizing the Euclidean distance between the fully digital and hybrid precoding matrices.

In addition, to address the performance degradation caused by low-resolution phase shifters, reference \cite{li2020dynamic} proposed a dynamic hybrid precoding scheme. In parallel, to reduce the hardware cost associated with conventional PSs, the studies in \cite{yan2022dynamic} introduced new architectures utilizing fixed PSs. 

The design methodologies for traditional antenna-based massive MIMO systems provide valuable insights and analytical frameworks that can inform the development of massive MIMO systems based on Rydberg sensors.

\subsection{Contributions}

Our major contributions are summarized as follows:
\begin{itemize}
	\item[$\bullet$]Based on recent Rydberg sensor advances, we classify array miniaturization techniques into two reuse strategies: LO-Shared (LO-S) versus LO-Dedicated (LO-D), and APD-Shared (APD-S) versus APD-Dedicated (APD-D). By analogy with conventional RF MIMO mapping, we introduce a reusable array model consisting of four fundamental architectures, D$\&$D, D$\&$S, S$\&$D, and S$\&$S, which represent all possible dedicated/shared combinations of LO and APD configurations.
	
	\item[$\bullet$] Based on the four fundamental architectures, we propose an optimization model for the hybrid Rydberg array combiner and solve it via alternating minimization of the Euclidean distance under both finite and infinite PS resolutions, with convergence analysis.
	
	\item[$\bullet$] We demonstrate that when LO reuse depth divides APD reuse depth, LO phase becomes irrelevant to performance due to full digital baseband compensation, enabling finite-resolution LO to achieve spectral efficiency equal to infinite-resolution systems.
	
	\item[$\bullet$] Under ideal precoding, we evaluate the four fundamental Rydberg array architectures and compare them with traditional RF arrays using partially-connected mapping, revealing reuse-related performance tradeoffs in both systems.
\end{itemize}

It should be noted that this work does not involve an analysis or comparison of the specific physical system of Rydberg atoms versus the response of traditional antennas. Therefore, the advantages or differences observed over conventional RF arrays in this study are attributed solely to the unique and inherent connection method of the Rydberg array antenna under sparse channel conditions, without presumed superiority or inferiority of the Rydberg atomic sensor in perceiving electromagnetic fields compared to traditional antennas.

\subsection{Organization}

The remainder of this paper is organized as follows. Section \ref{T2} presents the MIMO framework, design methodology, reuse architectures, and equivalent mapping strategies for Rydberg multiplexed arrays. Section \ref{T3} introduces the channel model and MIMO communication system model. Section \ref{T4} formulates the optimization problem for Rydberg array reuse architectures and develops an alternating minimization solution, with special attention to the case where LO reuse depth divides APD reuse depth, enabling a non-iterative direct solution with guaranteed performance equivalence. Section \ref{T5} provides numerical performance evaluation of the proposed configurations, and Section \ref{T6} concludes the paper.

\subsection{Notation}

The following notations are used throughout this paper. $a$, $\bm{a}$ and $\bm{A}$ stand for a scalar, column vector and matrix, respectively; $[\bm{A}]_{i,k}$ denotes the entry in the $i$-th row and $k$-th column of matrix $\bm{A}$, while $[\bm{A}]_{i:k,:}$ represents the submatrix of $\bm{A}$ consisting of rows $i$ through $k$ and all columns; The conjugate, transpose and conjugate transpose of $\bm{A}$ are represented by $\bm{A}^*$, $\bm{A}^T$ and $\bm{A}^H$; $\left| \bm{A}\right| $ and $\lVert \bm{A} \rVert_F$ denote the determinant and Frobenius norm of $\bm{A}$; $\otimes$ denotes the Kronecker product between matrices; $\bm{A}^{\dag}$ is the Moore-Penrose pseudo inverse of $\bm{A}$; Tr$\left[ \bm{A}\right] $ indicate the trace; $\mathbf{1}_{N_{lm}}$ denotes an all-ones column vector of dimension $N_{lm} \times 1$, and $\bm{0}$ represents an all-zero matrix of appropriate dimensions; $\text{arg}\left\{ a \right\} $ denotes the argument (phase angle) of $a$, while $|a|$ represents its magnitude (absolute value); $\bm{I}$ denotes the identity matrix; $\lceil a \rceil$ denotes the ceiling function, which returns the smallest integer greater than or equal to $a$.

%
%
%
%


\begin{figure*}
	\centering
	\includegraphics[width=0.95\textwidth]{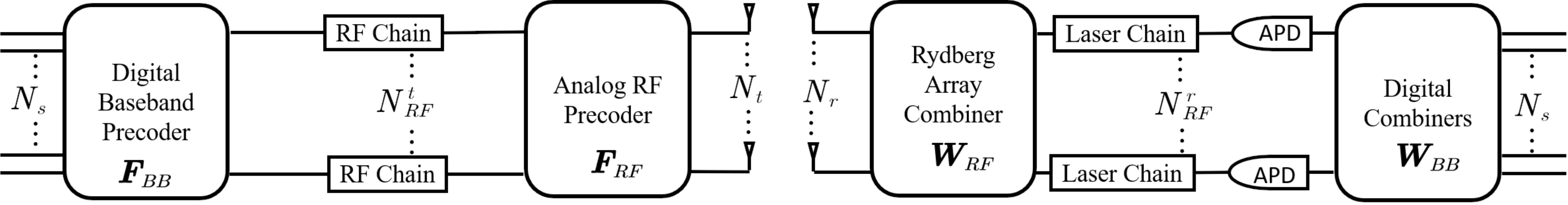}
	\caption{Hybrid architecture massive MIMO system block diagram with Rydberg atomic sensor implementation.}
	\label{architecture}
\end{figure*}

\begin{figure*}
	\centering
	\includegraphics[width=0.95\textwidth]{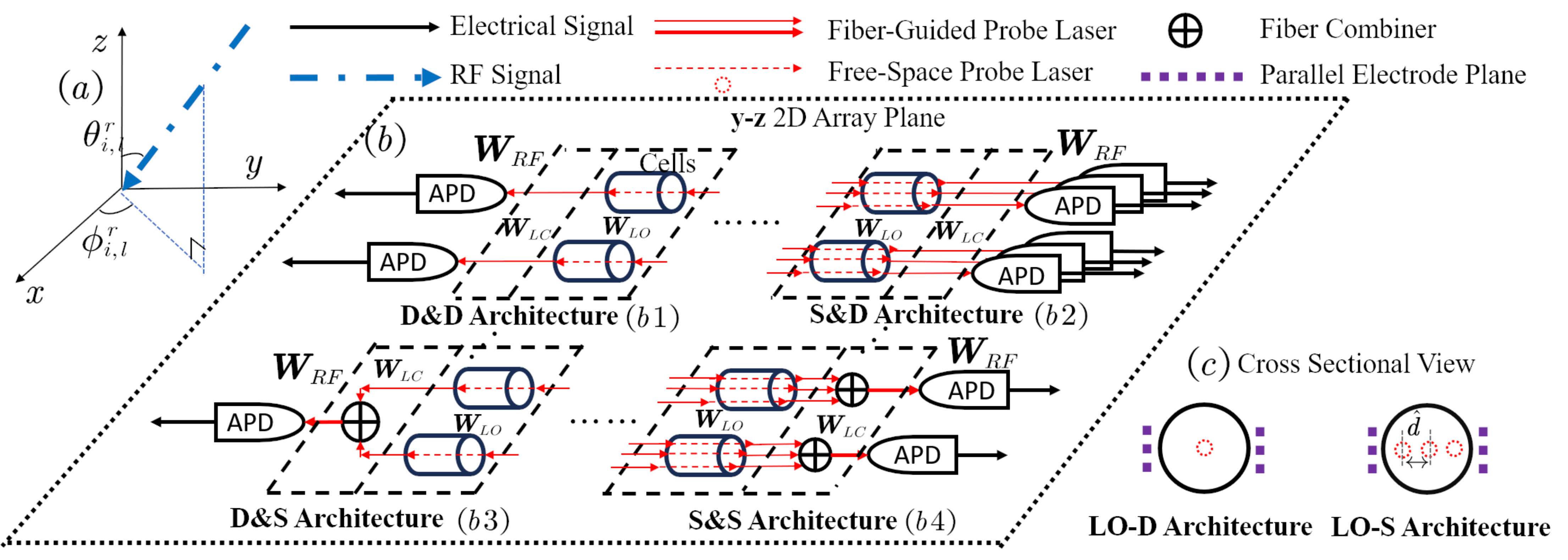}
	\caption{(a) The geometrical configuration of RF signal with azimuth angle $ \phi _{i,l}^{r}$ and elevation angle $\theta _{i,l}^{r}$. (b) The 3D view of four fundamental array configurations of Rydberg sensors$^{1}$. (c) The cross view of LO-S and LO-D architecture.}
	\label{ArrayModel}
\end{figure*}

\section{MIMO Hybrid Architecture For Rydberg Atomic Sensors}\label{T2}
We present single-user hybrid MIMO architecture for Rydberg sensor-based antennas, as illustrated in Fig. \ref{architecture}. In this architecture, a transmitter with $N_t$ traditional antennas sends $N_s$ data streams to a receiver equipped with $N_r$ Rydberg-based sensing elements. The system employs hybrid precoding, comprising a digital precoder $\bm{F}_{BB}$ and an analog precoder $\bm{F}_{RF}$ at the transmitter using traditional antenna array. At the receiver, following the same design principle, hybrid Rydberg antenna array is represented with an analog combiner $\bm{W}_{RF}$ (the Rydberg array combiner module) and a digital combiner module $\bm{W}_{BB}$, interconnected by optical chains that functionally correspond to RF chains in conventional hybrid architectures. Due to the analogous functionalities of RF chains and laser chains, their quantities are uniformly denoted as $N_{RF}^r$, subject to the constraints $N_s \leq N_{RF}^r \leq N_r$.

\subsection{The Model of Rydberg Array Combiner $\bm{W}_{RF}$}\label{T21}

The overall Rydberg array combining system $\bm{W}_{RF}$ consists of a low-dimensional digital combiner $\bm{W}_{BB}$ and a high-dimensional Rydberg array combiner $\bm{W}_{RF}$, with the key distinction that the latter is implemented using basic optical components rather than microwave devices. Such design rationale stems from the practical consideration that Rydberg-based reception typically requires a large number of APDs. Without a reuse architecture, where each Rydberg atomic vapor cell is assigned a dedicated APD and laser chain, the system would necessitate numerous APD and a complex optical setup, with each optical path requiring an individual LO signal. Such configuration represents the model adopted by most current Rydberg array implementations\cite{gong2025rydberg,zhu2025raq,atapattu2025detection,song2025csi,cui2025mimo,xu2025channel,cui2024multi,yuan2025electromagnetic,xu2024design}. By contrast, the introduction of shared optical components, such as laser/fiber combiners for APD reuse\cite{mao2022high,deb2018radio,simons2018fiber}, integrated local oscillators for LO reuse\cite{xie2025low,holloway2022rydberg}, and multiple laser pairs within a single atomic vapor cell for cell reuse\cite{wu2024enhancing,otto2021data}, enables significant simplification of the receiver structure and increases the number of antenna.

Based on these foundations, two reuse strategies, LO reuse and APD reuse, have been developed to simplify the array architecture. Building on existing Rydberg array design methodologies, we summarize four basic architectures, representing different combinations of dedicated and shared configurations for LO and APD, as illustrated in Fig. \ref{ArrayModel}(b)\footnote{Note that in Fig. \ref{ArrayModel}(b) and Fig. \ref{RYmapping}(d), the probe lasers served by one LO are combined into a group and received by a shared APD. In the array system considered in this work, we address the more general case where the number of probe lasers served by an LO can be fewer or greater than those combined for one APD, satisfying $N_r = N_{lon}\times N_{lom} = N_{lm} \times N_{RF}^r$.}

\begin{itemize}
	\item[$\bullet$] LO-Dedicated  $\&$ APD-Dedicated  (D$\&$D) Architecture, shown in Fig. \ref{ArrayModel}(b1).

	\item[$\bullet$] LO-Shared $\&$ APD-Dedicated  (S$\&$D) Architecture, shown in Fig. \ref{ArrayModel}(b2).
	
	\item[$\bullet$] LO-Dedicated $\&$ APD-Shared (D$\&$S) Architecture, shown in Fig. \ref{ArrayModel}(b3).
	
	\item[$\bullet$] LO-Shared $\&$ APD-Shared (S$\&$S) Architecture, shown in Fig. \ref{ArrayModel}(b4).
\end{itemize}

In Fig. \ref{ArrayModel}, we present a 3D view of a large-scale Rydberg array receiver based on mature and widely adopted fiber-optic components. \textit{All probe and coupling laser beams are fber-guided rather than via free-space propagation.} Within the atomic vapor cell, the laser beams are collimated using fiber collimators, and their separation is achieved through fiber-based wavelength division multiplexers. After transmission and separation, the probe laser remains in the optical fiber and is directly detected by APD. Furthermore, this work employs fiber combiners, which function mathematically as adders. Alternatively, following works \cite{otto2021data,wu2024enhancing}, lenses may be used to combine multiple laser beams in place of fiber combiners.

\begin{figure*}
	\centering
	\includegraphics[width=0.95\textwidth]{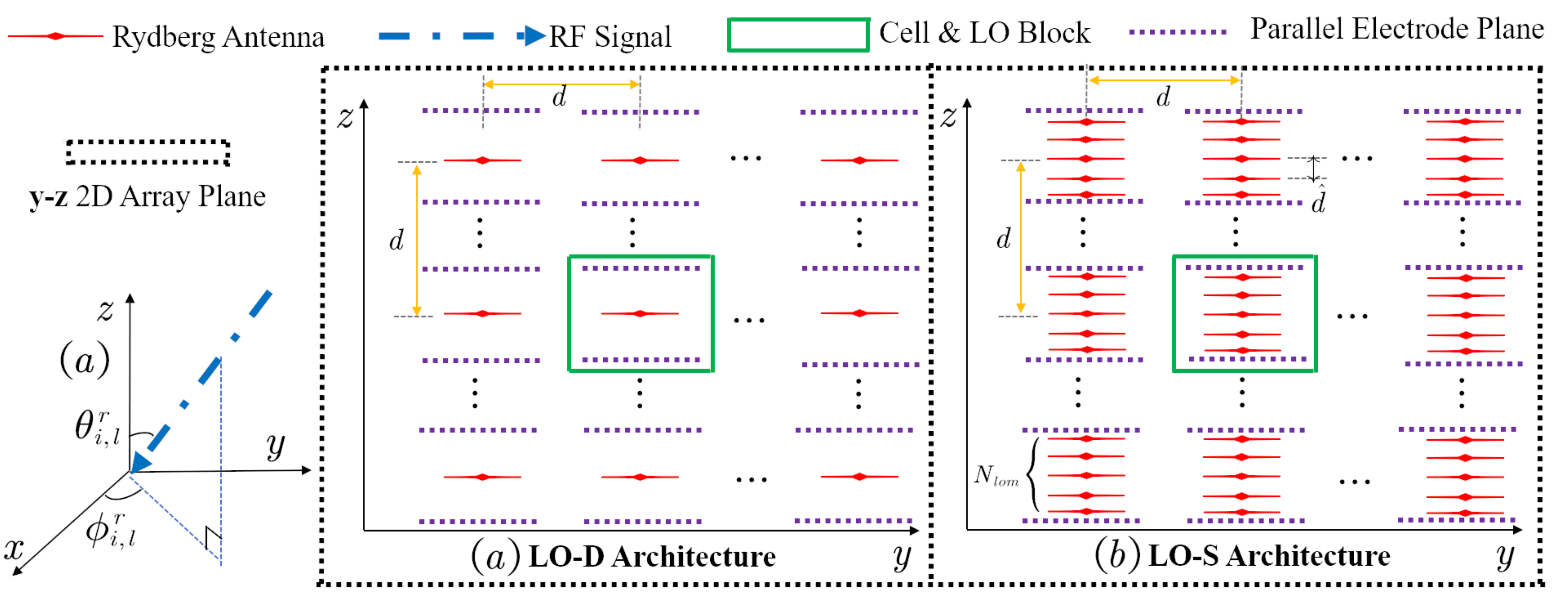}
	\caption{Rydberg 2D array plane. (a) The top view of LO-Dedicated architecture. (b) The top view of LO-Shared architecture.}
	\label{LOSDarchitecture}
\end{figure*}

It is worth noting that atomic vapor cell reuse refers to the configuration where a single atomic cell contains multiple laser pairs. When combined with LO reuse, all laser pairs within one atomic cell share a common local oscillator, which further enhances integration. In this paper, both atomic vapor cell reuse and local oscillator reuse are referred to as LO reuse and are illustrated as the fundamental building blocks, labeled as "Cell $\&$ LO Block", for both LO-D and LO-S architectures in Fig. \ref{LOSDarchitecture}. Although it is technically feasible to design one LO to support multiple atomic cells, such a configuration is functionally equivalent to the per-cell LO sharing approach. However, a drawback lies in that multiple atomic antennas (i.e., laser beams) share a single LO. Since the LO can be regarded as a phase shifter, the LO reuse design is equivalent to multiple antennas sharing a common phase shifter. In summary, LO reuse aims to maximize antenna density within constrained physical dimensions, while APD reuse seeks to minimize optical complexity and reduce the number of APD components.

Note that prior work in \cite{rotunno2023investigating} has demonstrated that the non-ideal characteristics of atomic vapor cells lead to non-uniform electric field distribution within the cell, resulting in inter-element coupling among atomic antennas in cell-reuse designs rather than their ideal independence. However, we posit that such effect stems primarily from material-level nonlinearities of the cell rather than the intrinsic properties of the four-level atomic physical system itself. Furthermore, it is anticipated that such non-uniformity can be mitigated through the use of high-transmissivity cell materials. Therefore, in this work, we assume ideal atomic vapor cells that maintain uniform internal field distribution. Under this assumption, the signals received by different antenna elements are mutually independent, and any relative phase difference between them arises solely from azimuth angle $\phi_{i,l}^{r}$ and elevation angle $\theta_{i,l}^{r}$ of the incident RF signal.

Based on the analysis and assumptions regarding the Rydberg receiver model, and considering that each laser chain, whether in a reused or non-reused configuration, connects on one side to optical signals and on the other to electrical signals converted by APD, we classify the signal processing stages accordingly. The part involved in subsequent electrical signal analysis is mathematically represented as $\bm{W}_{BB}$, while the part responsible for combining and processing optical signals is denoted as $\bm{W}_{RF}$. As shown in Fig. \ref{architecture}, $\bm{W}_{RF}$ block encompasses four basic configurations, visualized in the 3D schematic of Fig. \ref{ArrayModel}. The corresponding Rydberg 2D array layouts are presented in Fig. \ref{LOSDarchitecture}, while the equivalent mapping strategy for each configuration aligns with the connectivity patterns depicted in Fig. \ref{RYmapping}$^{1}$.

Specifically, the entire Rydberg array combiner module $\bm{W}_{RF}$ can be decomposed into two components, the local oscillator module and the laser connection (LC) module. This decomposition yields the mathematical representation is given by
\begin{equation}
	\begin{aligned}
\bm{W}_{RF}=\bm{W}_{LO}\bm{W}_{LC},
	\end{aligned}
\end{equation}
where $\bm{W}_{LO}$ denotes the equivalent mathematical matrix for either LO-Shared or LO-Dedicated array configurations, and $\bm{W}_{LC}$ represents the equivalent matrix for either APD-Shared or APD-Dedicated architectures.

\subsection{The Model of Laser Connection Module $\bm{W}_{LC}$}\label{T22}

To maintain minimal system complexity, this work deliberately employs simple fiber combiners, as shown in Fig. \ref{ArrayModel}(b3) and Fig. \ref{ArrayModel}(b4), which merge multiple laser beams into a single output and function mathematically as an adder, as the core optical component in the $\bm{W}_{LC}$ module. As a result, for APD-D architectures, $\bm{W}_{LC}=\bm{I}$ corresponds to a $N_r\times N_r$ identity matrix due to $N_r=N_{RF}^r$, while for APD-S architectures, it takes the form of a $N_r\times N_{RF}^{r}$ block diagonal matrix, as expressed below,
\begin{equation}
	\begin{aligned}
\boldsymbol{W}_{LC}=\left[ \begin{matrix}
	\mathbf{1}_{N_{lm}}&		\bm{0}&		\cdots&		\bm{0}\\
	\bm{0}&		\mathbf{1}_{N_{lm}}&		\cdots&		\bm{0}\\
	\vdots&		\vdots&		\ddots&		\vdots\\
	\bm{0}&		\bm{0}&		\cdots&		\mathbf{1}_{N_{lm}}\\
\end{matrix} \right] _{N_r\times N_{RF}^{r}},
	\end{aligned}
\end{equation}
where $N_{lm} = N_r / N_{RF}^{r}$ denotes the number of probe lasers combined by each fiber combiner. A total of $N_{RF}^{r}$ combiners, corresponding to laser chains and APDs (as shown by the fiber combiner in Fig. \ref{ArrayModel}(b) and the adder in Fig. \ref{RYmapping}), are used to merge the probe signals from all $N_r$ antennas. While more sophisticated optical devices could enable advanced functionalities, such as subarray-type configurations analogous to traditional RF antenna arrays\cite{li2020dynamic,yan2022dynamic} or optical components with phase-shifting capabilities, the essence of such approaches lies in modifying matrix $\bm{W}_{LC}$ by introducing complex exponential coefficients beyond simple 0 or 1 elements. However, these implementations would inevitably introduce significant hardware complexity and control overhead, which are not considered in this work.

\subsection{The Model of Local Oscillator Module $\bm{W}_{LO}$}\label{T23}

\begin{figure*}
	\centering
	\includegraphics[width=0.95\textwidth]{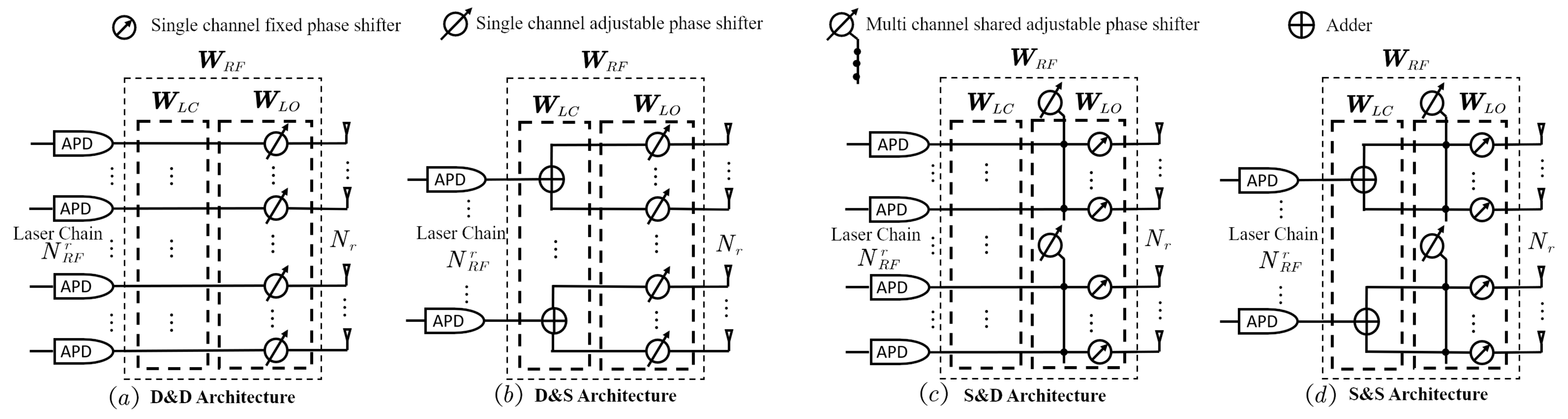}
	\caption{The equivalent mapping strategies for Rydberg atomic sensor-based massive MIMO system. (a) LO-Dedicated  $\&$ APD-Dedicated  (D$\&$D) Architecture. (b) LO-Dedicated  $\&$ APD-Shared  (D$\&$S) Architecture. (c) LO-Shared  $\&$ APD-Dedicated  (S$\&$D) Architecture. (d) LO-Shared  $\&$ APD-Shared  (S$\&$S) Architecture$^1$.}
	\label{RYmapping}
\end{figure*}

For LO-D architectures, each atomic antenna is equipped with an individual local oscillator source, i.e., the Cell $\&$ LO blocks in Fig. \ref{LOSDarchitecture}, which functions as a mixer, enabling independent phase control for every antenna element. Consequently, the resulting $\bm{W}_{LO}$ matrix for such architectures takes the form of a diagonal matrix, as expressed below,
\begin{equation}
	\begin{aligned}
\bm{W}_{LO}=\left[ \begin{matrix}
	e^{j\phi _1}&		0&		\cdots&		0\\
	0&		e^{j\phi _2}&		\cdots&		0\\
	\vdots&		\vdots&		\ddots&		0\\
	0&		0&		\cdots&		e^{j\phi _{N_r}}\\
\end{matrix} \right] _{N_r\times N_{r}},
	\end{aligned}
\end{equation}
where $\phi_{i}$ represents the phase shift introduced by the $i$-th local oscillator in LO-D architectures, with a total of $N_{lon} = N_r$ independent Cell $\&$ LO blocks serving as phase shifters for the $N_r=N_{lon}$ LO antennas (as shown by the cross sectional view in Fig. \ref{ArrayModel}(c), the 2D array plane in Fig. \ref{LOSDarchitecture}(a), and the single channel adjustable phase shifter in Fig. \ref{RYmapping}).

For LO-S architectures, multiple atomic antennas share a common local oscillator source, which inherently introduces correlated phase shifts across these antenna elements. Consequently, the resulting $\bm{W}_{LO}$ matrix for such architectures takes the form of a block diagonal matrix, as expressed below,
\begin{equation}
	\begin{aligned}
\boldsymbol{W}_{LO}=\left[ \begin{matrix}
	e^{j\phi _1}\boldsymbol{\tilde{I}}_1&		0&		\cdots&		0\\
	0&		e^{j\phi _2}\boldsymbol{\tilde{I}}_2&		\cdots&		0\\
	\vdots&		\vdots&		\ddots&		0\\
	0&		0&		\cdots&		e^{j\phi _{N_{lon}}}\boldsymbol{\tilde{I}}_{N_{lon}}\\
\end{matrix} \right] _{N_r\times N_r},
	\end{aligned}
\end{equation}
where $\phi_{i}$ represents the phase shift introduced by the $i$-th local oscillator in LO-S architectures, with a total of $N_{lon} $ independent Cell $\&$ LO blocks serving as phase shifters for the $N_r=N_{lon}\times N_{lom}$ LO antennas (as shown by the cross sectional view in Fig. \ref{ArrayModel}(c), the 2D array plane in Fig. \ref{LOSDarchitecture}(b), and the multi channel shared adjustable phase shifter in Fig. \ref{RYmapping}). $\boldsymbol{\tilde{I}}_{i}$ is a $N_{lom} \times N_{lom}$ diagonal matrix with complex exponential entries along its main diagonal, expressed as,
\begin{equation}
	\begin{aligned}
\boldsymbol{\tilde{I}}_i=\left[ \begin{matrix}
	e^{j\varphi _{i,1}}&		0&		\cdots&		0\\
	0&		e^{j\varphi _{i,2}}&		\cdots&		0\\
	\vdots&		\vdots&		\ddots&		0\\
	0&		0&		\cdots&		e^{j\varphi _{i,N_{lom}}}\\
\end{matrix} \right] _{N_{lom}\times N_{lom}},
	\end{aligned}
\end{equation}
where $N_{lom}=N_r/N_{lon}$ denotes the number of Rydberg atomic antennas (laser pairs) served by each Cell $\&$ LO block; $\varphi_{i,k}$ represents the relative phase difference among the multiple atomic antennas served by the same Cell $\&$ LO block. Such phase variation arises from different path lengths $\hat{d}$ between each atomic antenna and the shared local oscillator source, leading to distinct phase offsets in the received signals (as shown by the cross sectional view in Fig. \ref{ArrayModel}(c), the $\hat{d}$ in Fig. \ref{LOSDarchitecture}(b), and the single channel fixed phase shifter in Fig. \ref{RYmapping}). Given that the relative positions $\hat{d}$ of the laser pairs remain fixed in a configured system, we treat these phase values as constant parameters in our model. 


In summary, the combination of two $\bm{W}_{LO}$ matrix types, resulting from LO-S and LO-D architectures, with two $\bm{W}_{LC}$ matrix types, arising from APD-S and APD-D configurations, defines the four fundamental Rydberg array antenna architectures presented in this work. Note that the $\bm{W}_{RF}$ for all four fundamental architectures can be expressed as a product of two matrices, each being either diagonal or block diagonal. Therefore, they can be uniformly represented as
\begin{equation}
	\begin{aligned}
\boldsymbol{W}_{RF}=&\underset{\boldsymbol{W}_{LO}}{\underbrace{\left[ \begin{matrix}
			e^{j\phi _1}\boldsymbol{\tilde{I}}_1&		0&		\cdots&		0\\
			0&		e^{j\phi _2}\boldsymbol{\tilde{I}}_2&		\cdots&		0\\
			\vdots&		\vdots&		\ddots&		0\\
			0&		0&		\cdots&		e^{j\phi _{N_{lon}}}\boldsymbol{\tilde{I}}_{N_{lon}}\\
		\end{matrix} \right] }}
	\\
	&\ \ \ \ \ \ \ \ \ \ \ \ \ \ \ \  \times\underset{\boldsymbol{W}_{LC}}{\underbrace{\left[ \begin{matrix}
			\mathbf{1}_{N_{lm}}&		0&		\cdots&		0\\
			0&		\mathbf{1}_{N_{lm}}&		\cdots&		0\\
			\vdots&		\vdots&		\ddots&		\vdots\\
			0&		0&		\cdots&		\mathbf{1}_{N_{lm}}\\
		\end{matrix} \right] }},
	\end{aligned}
\end{equation}
where for $N_{lon} = N_r$ (i.e., $N_{lom} = 1$), we define $\boldsymbol{\tilde{I}}_{i} = 1$, indicating that the Rydberg array degenerates from an LO-S architecture to an LO-D architecture. Similarly, when $N_{ln} = 1$, the system degenerates from an APD-S architecture to an APD-D architecture.

It is noteworthy that although the Rydberg array in Fig. \ref{architecture} and Fig. \ref{ArrayModel} is equivalently represented as a network of phase shifters Fig. \ref{RYmapping}, these components do not physically exist in the actual system.

\section{Communication Channel and MIMO System Model}\label{T3} 
\subsection{Channel Model}\label{T31}

Typically, the operational frequency range of Rydberg atomic sensors spans from 1 GHz to 100 GHz, falling within the millimeter-wave spectrum. Therefore, the channel for Rydberg-based arrays can be appropriately characterized using the classical clustered model, such as the Saleh-Valenzuela model \cite{rappaport2015millimeter}. This model represents the channel matrix as having only a few dominant clusters of paths, given by

\begin{equation}
	\begin{aligned}
\bm{H}=&\sqrt{\frac{N_tN_r}{N_{cl}N_{ray}}}\sum_{i=1}^{N_{cl}}{\sum_{j=1}^{N_{ray}}{\alpha _{i,l}\bm{\Lambda }_r\left[ \phi _{i,l}^{r},\theta _{i,l}^{r} \right] \bm{\Lambda }_t\left[ \phi _{i,l}^{t},\theta _{i,l}^{t} \right]}}
\\
&\ \ \ \ \ \ \ \ \ \ \ \ \ \ \ \ \ \ \  \ \ \ \ \ \ \  \times \bm{a}_r\left[ \phi _{i,l}^{r},\theta _{i,l}^{r} \right] \bm{a}_t\left[ \phi _{i,l}^{t},\theta _{i,l}^{t} \right] ^H,
	\end{aligned}
\end{equation}
where the channel matrix $\bm{H}$ is constructed from $N_{{cl}}$ clusters, each comprising $N_{{ray}}$ paths. The gain of the $l$-th ray in the $i$-th cluster, denoted by $\alpha_{i,l}$, is an i.i.d. complex Gaussian random variable, i.e., $\alpha_{i,l} \sim \mathcal{CN}(0, \sigma_{\alpha, i}^2)$. The variances $\sigma_{\alpha, i}^2$ across clusters satisfy $\sum_{i=1}^{N_{{cl}}} \sigma_{\alpha, i}^2 = \hat{\gamma}$, a normalization factor that ensures $\mathbb{E}\left[\lVert \bm{H} \rVert_F^2 \right] = N_t N_r$. Furthermore, $\bm{a}_r\left[ \phi _{i,l}^{r},\theta _{i,l}^{r} \right] $ and $\bm{a}_t\left[ \phi _{i,l}^{t},\theta _{i,l}^{t} \right] $ represent the receive and transmit array response vectors, respectively, where $ \phi _{i,l}^{r}$, $\phi _{i,l}^{t}$ and $\theta _{i,l}^{r}$, $\theta _{i,l}^{t}$ denote the azimuth and elevation angles of arrival (AoAs) and departure (AoDs).

For a square uniform planar array (UPA) comprising $N_{}$ antenna elements arranged in a $\sqrt{N_{}} \times \sqrt{N_{}}$ grid, the array response vector for the $l$-th ray in the $i$-th cluster is given by
\begin{equation}
	\begin{aligned}
\bm{a}\left[ \phi _{i,l},\theta _{i,l} \right]=\frac{1}{\sqrt{N_{}}}\left[ 1,\cdots ,e^{j\frac{2\pi}{\lambda}d\left( q_1\sin \phi _{i,l}\sin \theta _{i,l}+q_2\cos \theta _{i,l} \right)}, \right. 
\\
\left. \cdots ,e^{j\frac{2\pi}{\lambda}d\left( \left( \sqrt{N_{}}-1 \right) \sin \phi _{i,l}\sin \theta _{i,l}+\left( \sqrt{N_{}}-1 \right) \cos \theta _{i,l} \right)} \right] ^T,
	\end{aligned}
\end{equation}
where $d$ and $\lambda$ are the antenna spaceing and signal wavelength; $0 \le  q_1 \le \sqrt{N}-1$ and $0 \le  q_2 \le \sqrt{N}-1$ are the antenna indices in the 2D plane. In our analysis, the transmitting antenna is considered, without loss of generality, to be of this fundamental type, i.e., $\bm{a}_t\left[ \phi _{i,l}^{t},\theta _{i,l}^{t} \right]=\bm{a}\left[ \phi _{i,l}^{t},\theta _{i,l}^{t} \right]$ and $N =N_t$.

For the receiving Rydberg array, the antenna response vector differs between LO-S and LO-D architectures. Although each Rydberg antenna (laser pair) physically exhibits a cylindrical geometry, we follow common practice in the literature by modeling them as point antennas with uniform inter-element spacing $d$ between adjacent Cell $\&$ LO block centers. Top-view schematics of both antenna configurations are presented in Fig. \ref{LOSDarchitecture}. 

Given that the physical size of a Rydberg array antenna is constrained by the dimensions and number $N_{lon}$ of atomic vapor cells and LO blocks, and considering the advantage of LO-S over LO-D in accommodating more antenna elements within a fixed area (i.e., for identical $N_{lon}$), we model the array response vector of the LO-D architecture as that of a classical uniform square planar array, and accordingly develop the LO-S array response model under the same aperture size for comparing LO-S and LO-D architectures.

Thus, the LO-D architecture shares quite similar structure as conventional RF antennas, forming a standard uniform square planar array with the array response vector given by

\begin{equation}
	\begin{aligned}
		\bm{a}_{r}^{D}\left[ \phi _{i,l}^{t},\theta _{i,l}^{t} \right] =\bm{a}\left[ \phi _{i,l}^{t},\theta _{i,l}^{t} \right], 
	\end{aligned}
\end{equation}
where the top-view schematics of LO-D configuration and its corresponding Cell $\&$ LO block are presented in Fig. \ref{LOSDarchitecture}(a). In summary, for the LO-D architecture, each Rydberg array consists of $N_{lon}$ Cell $\&$ LO blocks, with each block containing a single antenna element. For a $N_r$ element array, the total area is given by $S = N_{lon} d^2 = N_r d^2$, and $\bm{a}_{r}^{D}\left[ \phi _{i,l}^{t},\theta _{i,l}^{t} \right]$ is a $N_{lon} \times 1$ vector.

In the LO-S configuration, the array forms non-uniform planar array (Non-UPA) as antenna elements, and are densely integrated along the z-axis with increased count, while the number of blocks $N_{lon}$ and the physical area remain fixed at $S = N_{lon} d^2$. The corresponding antenna response vector can therefore be expressed as
\begin{equation}
	\begin{aligned}
\bm{a}_{r}^{S}\left[ \phi _{i,l}^{t},\theta _{i,l}^{t} \right] =\bm{a}\left[ \phi _{i,l}^{t},\theta _{i,l}^{t} \right] \otimes \bm{a}_z\left[ \theta _{i,l}^{t} \right], 
	\end{aligned}
\end{equation}
where $\bm{a}\left[ \phi _{i,l}^{t},\theta _{i,l}^{t} \right]$ is a $N_{lon} \times 1$ vector, $\bm{a}_{r}^{S}\left[ \phi _{i,l}^{t},\theta _{i,l}^{t} \right]$ is a $N_{lon}N_{lom} \times 1$ vector, and $\bm{a}_z\left[ \theta _{i,l}^{t} \right]$ is given by
\begin{equation}
	\begin{aligned}
\bm{a}_z\left[ \theta _{i,l}^{t} \right] =\frac{1}{\sqrt{N_{lom}}}\left[ \begin{array}{c}
	e^{j\frac{2\pi}{\lambda}\left( 0-\frac{N_{lom}-1}{2} \right) \hat{d}\cos \theta _{i,l}^{t}}\\
	e^{j\frac{2\pi}{\lambda}\left( 1-\frac{N_{lom}-1}{2} \right) \hat{d}\cos \theta _{i,l}^{t}}\\
	\vdots\\
	e^{j\frac{2\pi}{\lambda}\left( N_{lom}-1-\frac{N_{lom}-1}{2} \right) \hat{d}\cos \theta _{i,l}^{t}}\\
\end{array} \right] ,
	\end{aligned}
\end{equation}
where the equivalent top-view layout is shown in Fig. \ref{LOSDarchitecture}(b). 

Note that in practice each atomic vapor cell is cylindrical and optical components occupy physical space, making it challenging to ensure that every antenna block is perfectly square with uniform spacing $d$ and internal spacing $\hat{d}$. However, both $d$ and $\hat{d}$ remain fixed for a given array configuration and can be precisely measured. Furthermore, the specific values of $d$ and $\hat{d}$ do not affect the sparsity of the channel. Therefore, without loss of generality, we assume uniform inter-block spacing $d$ and intra-block spacing $\hat{d}=0.1d$ throughout the array. Importantly, the array response vector formulation and subsequent analytical methodology remain fully applicable even with non-uniform $d$ and $\hat{d}$ spacing.

In addition, function $\bm{\Lambda }_r\left[ \phi _{i,l}^{r},\theta _{i,l}^{r} \right]$ and  $ \bm{\Lambda }_t\left[ \phi _{i,l}^{t},\theta _{i,l}^{t} \right]$ represent the transmit and receive antenna element gain at the corresponding angles of departure and arrival. As for the conventional RF transmitter, several parametric mathematical models have been proposed for the function $ \boldsymbol{\Lambda }_t\left[ \phi _{i,l}^{t},\theta _{i,l}^{t} \right]$. For instance, adopting the ideal sectored element model for the transmitter antennas\cite{singh2011interference}, function  $ \bm{\Lambda }_t\left[ \phi _{i,l}^{t},\theta _{i,l}^{t} \right]$ is given by
\begin{equation}\label{AEG}
	\begin{aligned}
		&\mathbf{\Lambda }_t\left[ \phi _{i,l}^{t},\theta _{i,l}^{t} \right]=1,
	\end{aligned}
\end{equation}
where the antenna gain is isotropic with respect to direction $\phi _{i,l}^{t}$ and $\theta_{i,l}^{t}$. Alternatively to the simplified model, the functions $\mathbf{\Lambda }_t\left[ \phi _{i,l}^{t},\theta _{i,l}^{t} \right]$ can be substituted with empirical far-field radiation patterns of conventional antennas, such as patch or half-wave dipole antennas\cite{balanis2016antenna}.

For the receiving Rydberg array antenna, no theoretical or empirical model currently establishes the relationship between its unit antenna element gain and the azimuth/elevation angles. The most relevant work in this context \cite{robinson2021determining} addresses related aspects but still does not provide a complete model characterizing this fundamental relationship. Consequently, we adopt the same simplified model presented in Eq. (\ref{AEG}). An essential clarification must be made, since each antenna element operates independently without mutual coupling, any deviation of the actual Rydberg unit antenna element gain $\bm{\Lambda }_r\left[ \phi _{i,l}^{r},\theta _{i,l}^{r} \right]$ from the ideal sectored element model in Eq. (\ref{AEG}) would alter channel matrix $\bm{H}$ but preserve its sparsity. Such sparsity is maintained due to the inherent properties of the mmWave channel, characterized by $N_{cl}$ scattering clusters and $N_{ray}$ propagation paths. This rationale allows us to incorporate the angular dependence of antenna gain into the statistical variation of $\bm{H}$, thereby ensuring that the uncertainty in the actual antenna element gain function $\mathbf{\Lambda }_r\left[ \phi _{i,l}^{r},\theta _{i,l}^{r} \right]$, including those arising from the length of atomic vapor cell as examined in \cite{guo2025aoa} for non-point sensor behavior, does not affect the validity of our subsequent analysis.

\subsection{MIMO System Model}\label{T32}

The transmitted signal is given by $\bm{x} = \bm{F}_{RF} \bm{F}_{BB} \bm{s}$, where $\bm{s}$ is the $N_s \times 1$ symbol vector satisfying $\mathbb{E}[\bm{s}\bm{s}^H] = \frac{1}{N_s} \bm{I}_{N_s}$. Here, $\bm{F}_{BB}$ is the digital baseband precoder, and $\bm{F}_{RF}$ is the analog RF precoder, which together are subject to the normalized power constraint $||\bm{F}_{RF}\bm{F}_{BB}||_F^2 = N_s$. Assuming a narrow-band block-fading channel, the processed received signal is expressed as
\begin{equation}
	\begin{aligned}
\bm{y}=\sqrt{\rho}\boldsymbol{W}_{BB}^{H}\boldsymbol{W}_{RF}^{H}\bm{H}\boldsymbol{F}_{RF}\boldsymbol{F}_{BB}\bm{s}+\boldsymbol{W}_{BB}^{H}\boldsymbol{W}_{RF}^{H}\bm{n},
	\end{aligned}
\end{equation}
where $\rho$ represents the average received power, $\bm{H}$ is the channel matrix, and $\bm{n} \sim \mathcal{CN}(\mathbf{0}, \sigma_n^2 \bm{I})$ is the additive noise vector. The receiver employs a hybrid combiner, composed of an analog RF combiner $\bm{W}_{RF}$ and a digital baseband combiner $\bm{W}_{BB}$. We assume perfect channel state information (CSI) is available at both ends, which is a reasonable assumption given that CSI can be acquired at the receiver via channel estimation \cite{alkhateeb2014channel} and subsequently fed back to the transmitter using efficient techniques \cite{el2014spatially,love2005limited}. Assuming Gaussian signaling, the achievable spectral efficiency is given by
\begin{equation}
	\begin{aligned}
R=&\log _2\left| \boldsymbol{I}_{N_s}+\frac{\rho}{\sigma _{n}^{2}N_s}\left( \boldsymbol{W}_{RF}\boldsymbol{W}_{BB} \right) ^{\dag}\bm{H}\boldsymbol{F}_{RF}\boldsymbol{F}_{BB} \right. 
\\
&\ \ \ \ \ \ \ \ \ \ \ \ \ \ \ \ \ \ \ \ \ \ \ \ \ \ \ \ \ \ \left. \times \boldsymbol{F}_{BB}^{H}\boldsymbol{F}_{RF}^{H}\bm{H}^H\boldsymbol{W}_{RF}\boldsymbol{W}_{BB} \right|.
	\end{aligned}
\end{equation}

Furthermore, this work does not address specific Rydberg physical models. Any observed advantages or distinctions compared to conventional RF arrays stem from the unique connectivity of the Rydberg array under sparse channel conditions and low-complexity array design, rather than from fundamental electric field detection properties of the Rydberg atomic system itself. This is because under conditions where the local oscillator intensity significantly exceeds the signal strength and superheterodyne reception is employed, the detection scheme essential for current Rydberg-based high-sensitivity measurements rather than splitting-based approaches, the Rydberg sensor exhibits general linearity and functions as a microwave receiving component indistinguishable from conventional antennas. Alternatively, under specific modulation schemes and power levels as demonstrated in work \cite{wu2025analysis}, the nonlinearity exhibited by Rydberg sensors can be equivalently modeled as additive white Gaussian noise. Thus, for comparative purposes, the Rydberg array can be considered linear and exhibits lower noise $\bm{n}$ and $\sigma_n^2$ than conventional antennas.

\section{Problem Description and Analysis}\label{T4} 
\subsection{Problem Statement}\label{T41}

As established in \cite{lee2014hybrid,yu2016alternating,el2014spatially,chen2017hybrid,lyu2021lattice}, the joint design of precoders and decoders can be decoupled into two separate optimization subproblems, given by

\begin{equation}
	\begin{aligned}
&\min_{\boldsymbol{F}_{BB},\boldsymbol{F}_{RF}}\lVert \boldsymbol{F}_{{opt}}-\boldsymbol{F}_{RF}\boldsymbol{F}_{BB} \rVert _F\quad 
\\
&\ \ \ \ \text{s.t.\quad }\boldsymbol{F}_{RF}\in \mathcal{F}
\\
&\ \ \ \ \ \ \ \  \ \ \lVert \boldsymbol{F}_{RF}\boldsymbol{F}_{BB} \rVert _{F}^{2}=N_s,
	\end{aligned}
\end{equation}
and 
\begin{equation}
	\begin{aligned}
&\min_{\bm{W}_{BB},\bm{W}_{RF}}\lVert \bm{W}_{{opt}}-\bm{W}_{RF}\bm{W}_{BB} \rVert _F\quad 
\\
&\ \ \ \ \ \text{s.t.\quad }\bm{W}_{RF}\in \mathcal{F},
	\end{aligned}
\end{equation}
where $\bm{F}_{opt}$ and $\bm{W}_{opt}$ are the optimal fully digital precoder and combiner, defined as the first $N_s$ columns of the right-singular matrix $\bm{V}$ and left-singular matrix $\bm{U}$ from the SVD $\bm{H} = \bm{U}\bm{\Sigma}\bm{V}^H$, respectively, with $\bm{\Sigma}$ having diagonals sorted in descending order. $\mathcal{F}$ denotes the feasible set for the local oscillator phase shifts $\phi_i$ in Sec. \ref{T23}, which may encompass either finite or infinite resolution configurations.

While various methods have been developed to address this classical optimization problem at the transmitter, this work simplifies the system model to align with its core focus, receiver-side analysis. Specifically, we assume an ideal transmitter without power constraints and perfect precoding, defined as $\bm{F}_{opt} = \bm{F}_{RF}\bm{F}_{BB}$, and concentrate solely on the optimization of the receiver combiner, given by
\begin{equation}\label{CoreProblem}
	\begin{aligned}
		&\min_{\bm{W}_{BB},\bm{W}_{LO}}\lVert \bm{W}_{{opt}}-\bm{W}_{LO}\bm{W}_{LC}\bm{W}_{BB} \rVert _F\quad 
		\\
		&\ \ \ \ \ \ \ \ \ \text{s.t.\quad }\bm{W}_{LO}\in \mathcal{F},
	\end{aligned}
\end{equation}
where $\bm{W}_{LC}$ is a fixed matrix, determined solely by the choice of APD-S or APD-D architecture, requiring optimization only of $\bm{W}_{LO}$ and $\bm{W}_{BB}$. 

\subsection{Solution Methodology}

Classically, the proposed optimization problem is tackled through the use of a two-stage iterative procedure based on the alternating minimization technique\cite{yu2016alternating} to find $\bm{W}_{LO}$ and $\bm{W}_{BB}$. Specifically, the algorithm alternates between fixing $\bm{W}_{LO}$ to optimize $\bm{W}_{BB}$, and fixing $\bm{W}_{BB}$ to optimize $\bm{W}_{LO}$, repeating this process until the change after $p$ iterations falls below a predefined threshold $\varepsilon$, given by
\begin{equation}\label{threshold}
	\begin{aligned}
&\left| \lVert \boldsymbol{W}_{opt}-\boldsymbol{W}_{LO}^{\left[ p+1 \right]}\boldsymbol{W}_{LC}\boldsymbol{W}_{BB}^{\left[ p \right]} \rVert _{F}^{2} \right. 
\\
&\ \ \ \ \ \ \ \ \ \ \ \ \ \ \ \ \ \left. -\lVert \boldsymbol{W}_{opt}-\boldsymbol{W}_{LO}^{\left[ p+2 \right]}\boldsymbol{W}_{LC}\boldsymbol{W}_{BB}^{\left[ p+1 \right]} \rVert _{F}^{2} \right|<\varepsilon .
	\end{aligned}
\end{equation}

\subsubsection{$\bm{W}_{BB}$ Optimization with Fixed $\bm{W}_{LO}$} The problem in Eq. (\ref{CoreProblem}) becomes the linear squares problem with unique solution given by,
\begin{equation}\label{step1}
	\begin{aligned}
\boldsymbol{W}_{BB}=\left( \boldsymbol{W}_{LO}\boldsymbol{W}_{LC} \right) ^{\dag}\boldsymbol{W}_{opt}.
	\end{aligned}
\end{equation}

\subsubsection{$\bm{W}_{LO}$ Optimization with Fixed $\bm{W}_{BB}$} Since $\bm{W}_{LO}$ is either a block diagonal matrix (for LO-S architectures) or a diagonal matrix (for LO-D architectures), the problem in Eq. (\ref{CoreProblem}) can be simplified and reformulated as

\begin{equation}\label{SubProblem2_1}
	\begin{aligned}
		&\min_{\boldsymbol{W}_{LO}} \left\| \boldsymbol{W}_{opt} - \boldsymbol{W}_{LO} \boldsymbol{W}_{LC} \boldsymbol{W}_{BB} \right\|_F^2 \\
	 =& \min_{\{\phi_i\}} \sum_{i=1}^{N_{lon}} 
		\left\| [\boldsymbol{W}_{opt}]_{(i-1)N_{lom}+1:iN_{lom},:} \right. \\
		&\qquad \qquad  \left. - e^{j\phi_i} \boldsymbol{\tilde{I}}_i [\boldsymbol{W}_{LC} \boldsymbol{W}_{BB}]_{(i-1)N_{lom}+1:iN_{lom},:} \right\|_F^2,
	\end{aligned}
\end{equation}
where the original $\boldsymbol{W}_{LO}$ optimization problem can be decomposed into $N_{lon}$ independent subproblems. For each subproblem, this is essentially a matrix approximation problem using phase rotation, and a closed-form expression exists for each $\phi_i$, given by

\begin{equation}
	\begin{aligned}\label{IWLO}
\phi _i&=\text{arg}\left\{ \text{Tr}\left[ \left[ \boldsymbol{W}_{LC}\boldsymbol{W}_{BB} \right] _{\left( i-1 \right) N_{lom}+1:iN_{lom},:}^{H}\boldsymbol{\tilde{I}}_{i}^{H} \right. \right.\\
&\ \ \ \ \ \ \ \ \ \ \ \ \ \ \ \ \ \ \ \ \left. \left. \times \left[ \boldsymbol{W}_{opt} \right] _{\left( i-1 \right) N_{lom}+1:iN_{lom},:} \right] \right\} ,
	\end{aligned}
\end{equation}
where the detailed derivation is provided in Appendix \ref{A11}.

When considering finite resolution local oscillator sources, analogous to the deployment of finite resolution phase shifters in traditional fully-connected architectures, the limited phase combinations result in a finite set of possible $\bm{W}_{RF}$ configurations. Given $B$ bits of resolution, the feasible set of phases is given by
\begin{equation}
	\begin{aligned}
\mathcal{B}=\left\{ e^{j\times 0},e^{j\times \left( 2\pi /2^B \right)},\cdots ,e^{j\times \left( 2\pi /2^B \right) \times \left( 2^B-1 \right)} \right\} .
	\end{aligned}
\end{equation}

 Consequently, without considering combinatorial complexity, an exhaustive search algorithm can be used to find the global optimum, for the problem and solution formulated in Eq. (\ref{SubProblem2_1}) and Eq. (\ref{IWLO}), since the obtained $\phi_i$ may not satisfy $\phi_i \in \mathcal{B}$, an additional quantization step is required, given by
\begin{equation}\label{Q}
	\begin{aligned}
		\hat{\phi}_i=\mathcal{Q}\left[ \phi _i \right] =\frac{2\pi}{2^B}\hat{b} ,
	\end{aligned}
\end{equation}
where $\hat{\phi}_i$ is the quantized version of $\phi _i$. $\mathcal{Q}\left[ \cdot \right]$ is a quantizer that quantizes its input to the nearest point in the set $\mathcal{B}$, given by
\begin{equation}
	\begin{aligned}
		\hat{b} = \arg\min_{b \in \{0,1,\cdots,2^B-1\}} \left| \phi_i - \frac{2\pi}{2^B}b \right|.
	\end{aligned}
\end{equation}

Since $\bm{W}_{LO}$ is diagonal, the optimal $\bm{W}_{LO}^{[p]}$ corresponding to each iteration's $\bm{W}_{BB}^{[p]}$ can be obtained by simply quantizing the value of each optimized phase shift to a finite set $\mathcal{B}$, without requiring complex quantization procedures or extensive computation. The detailed analysis is provided in Appendix \ref{A12}.

\begin{figure}
	\centering
	\includegraphics[width=0.42\textwidth]{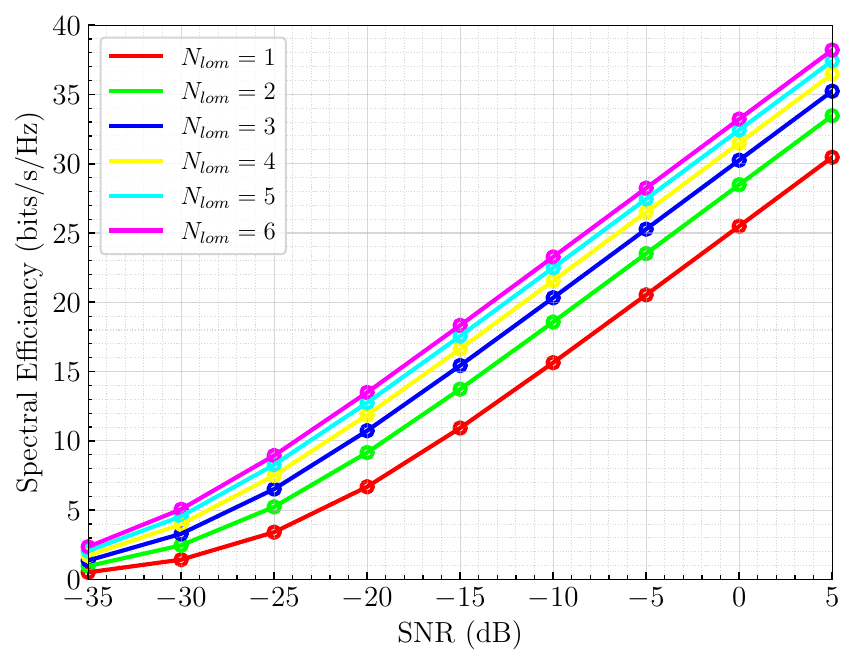}
	\caption{Spectral efficiency under $N_s=3$.}
	\label{LO_DS_Ns3}
\end{figure}

\begin{algorithm}
	\caption{}
	\label{alternating_minimization}
	\begin{algorithmic}[1]
		\Require $\bm{W}_{opt}$ and $\bm{W}_{LC}$;
		\State Initialize: $\boldsymbol{W}_{LO}^{\left[1 \right] }$, $\boldsymbol{W}_{BB}^{\left[1 \right] }$, $p \leftarrow 1$;
		\Repeat
		\State $\bm{W}_{BB}^{\left[p+1 \right] }$ according to Eq. (\ref{step1});
		\State Obtain $\boldsymbol{W}_{LO}^{\left[p+2 \right] }$ according to Eq. (\ref{IWLO}) and Eq. (\ref{Q});
		\State $p \leftarrow p + 1$;
		\Until{A stopping criterion triggers arrcoding to Eq. (\ref{threshold})};
		\State Output: $\bm{W}_{BB}^{\left[p \right] }$, $\bm{W}_{LO}^{\left[p+1 \right] }$.
	\end{algorithmic}
\end{algorithm}

\subsubsection{Convergence Analysis}For the two-step alternating optimization strategy shown in Algorithm \ref{alternating_minimization}, for fixed $\bm{W}_{LO}$, the globally optimal $\bm{W}_{BB}$ is given by the solution in Eq. (\ref{step1}). For fixed $\bm{W}_{BB}$, as shown in the appendix \ref{A1}, the solution for $\bm{W}_{LO}$ derived from Eq. (\ref{IWLO}) and Eq. (\ref{Q}) are globally optimal for the finite and infinite phase resolution, respectively. Therefore, for the objective function $g\left[ \boldsymbol{W}_{BB},\boldsymbol{W}_{LO} \right] =\lVert \boldsymbol{W}_{opt}-\boldsymbol{W}_{LO}\boldsymbol{W}_{LC}\boldsymbol{W}_{BB} \rVert _F$, the $\left( p+1\right) $-th iteration satisfies

\begin{equation}
	\begin{aligned}
		g\left[ \boldsymbol{W}_{BB}^{\left[ p \right]},\boldsymbol{W}_{LO}^{\left[ p \right]} \right] \ge g\left[ \boldsymbol{W}_{BB}^{\left[ p+1 \right]},\boldsymbol{W}_{LO}^{\left[ p \right]} \right] \ge g\left[ \boldsymbol{W}_{BB}^{\left[ p+1 \right]},\boldsymbol{W}_{LO}^{\left[ p+1 \right]} \right] ,
	\end{aligned}
\end{equation}
where the optimality of each step in the alternating optimization process ensures convergence.

\begin{figure*}
	\centering
	\subfigure[]{
		\begin{minipage}[t]{0.5\linewidth}
			\centering
			\includegraphics[width=3.1 in]{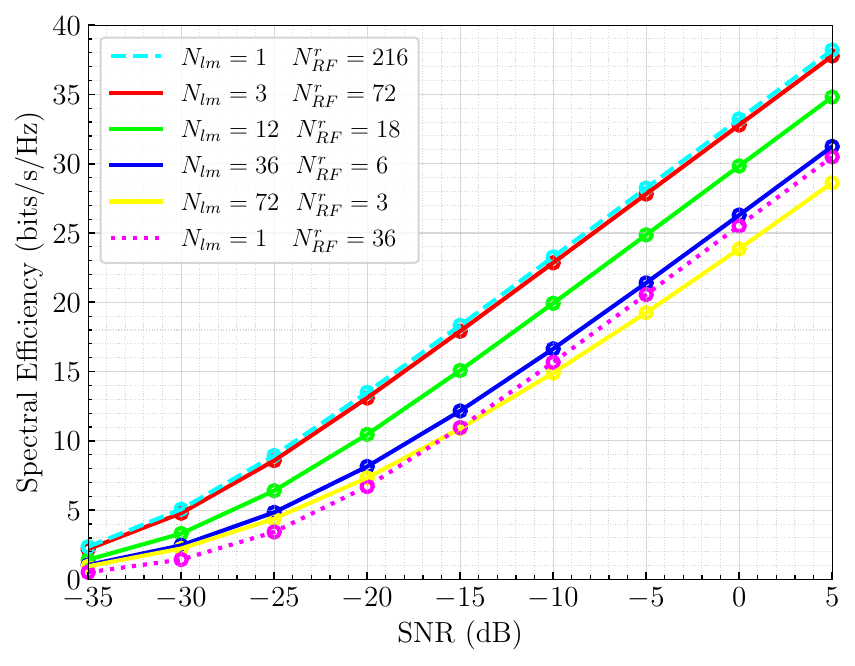}
			\label{LC_DS_Ns3_N_lom6}
		\end{minipage}%
	}%
	\subfigure[]{
		\begin{minipage}[t]{0.5\linewidth}
			\centering
			\includegraphics[width=3.1in]{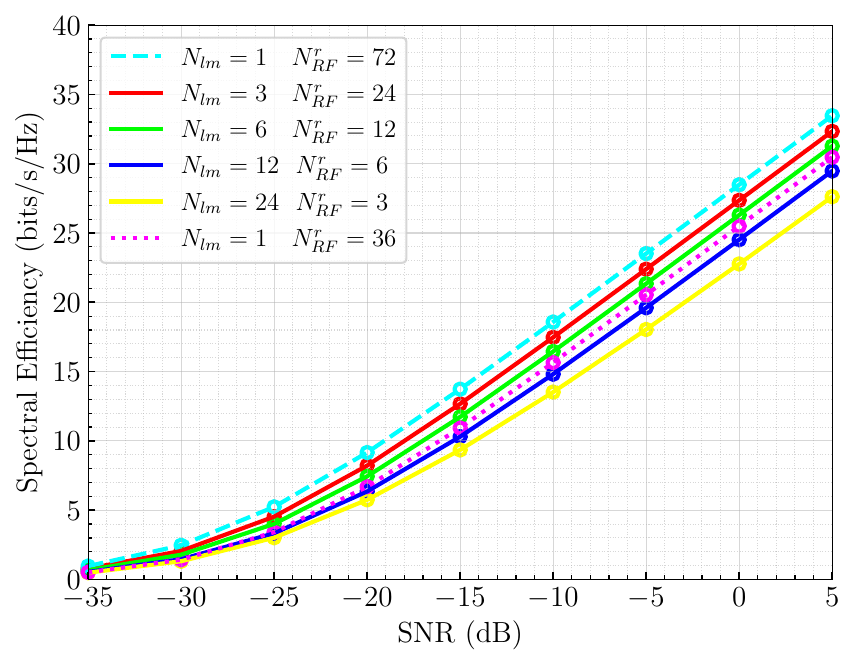}
			\label{PYLC_DS_Ns3_N_lom2}
		\end{minipage}
	}%
	\centering
	\caption{Spectral efficiency performance with respect to $N_{RF}^r$ and $N_{lom}$. (a) $N_s=3$ and $N_r=N_{lon}\times N_{lom} =36\times 6$. (c) $N_s=3$ and $N_r=N_{lon}\times N_{lom} =36\times 2$.}
	\label{LCReuse}
\end{figure*}

\subsection{Typical Scenario Analysis: Proportional LO and APD Reuse Configuration}\label{T34}

As a straightforward special case, in the LO-D architecture without APD reuse (i.e., when $\bm{W}_{LC} = \bm{I}$), and since the received signal in a Rydberg-based system is automatically down-converted to baseband, no explicit LO phase adjustment is required. All received antenna signals are directly converted to baseband, making the system equivalent to a fully digital MIMO receiver. Therefore, finite LO resolution does not compromise performance. In this configuration, the local oscillator serves only to extract the phase of the received signal. 

This phenomenon also occurs when the LO reuse depth $N_{lom}$ divides the APD reuse depth $N_{lm}$, resulting in an effect analogous to that of fully digital combiners, which we term reuse equivalent fully digital combiners. In such cases, the iterative procedure in Algorithm \ref{alternating_minimization} is no longer required, enabling direct optimization of $\bm{W}_{LO}$ and $\bm{W}_{BB}$ (or $\bm{\hat{W}}_{LO}$ and $\bm{\hat{W}}_{BB}$).

\subsubsection{Typical Scenario Description and Analysis}
When APD reuse is implemented and $N_{lm}$ is divisible by $N_{lom}$, APD reuse occurs exclusively within each Cell $\&$ LO block, with no sharing across different blocks. In this case, $\boldsymbol{\tilde{I}}_i$ can be expressed as
\begin{equation}
\begin{aligned}
\boldsymbol{\tilde{I}}_i=\left[ \begin{matrix}
		\boldsymbol{\mathring{I}}_{i,1}&		\mathbf{0}&		\cdots&		\mathbf{0}\\
		\mathbf{0}&		\boldsymbol{\mathring{I}}_{i,2}&		\cdots&		\mathbf{0}\\
		\vdots&		\vdots&		\ddots&		\mathbf{0}\\
		\mathbf{0}&		\mathbf{0}&		\cdots&		\boldsymbol{\mathring{I}}_{i,N_{lom}/N_{lm}}\\
	\end{matrix} \right] _{N_{lom}\times N_{lom}},
		\end{aligned}
\end{equation}
where the $N_{lom}$ diagonal elements $\varphi_{i,k}$ can be partitioned into $N_{lom}/N_{lm}$ smaller diagonal matrices, denoted $\boldsymbol{\mathring{I}}_{i,1}$ to $\boldsymbol{\mathring{I}}_{i,N_{lom}/N_{lm}}$. Then, $\bm{W}_{RF}$ can be equivalently expressed as the corresponding product of these block matrices from $\bm{W}_{LO}$ and $\bm{W}_{LC}$, represented explicitly as a block diagonal matrix in Eq. (\ref{DEQ}), where each $\phi_i$ is multiplied by $N_{lom}/N_{lm}$ matrices, specifically from $\boldsymbol{\mathring{I}}_{i,1}$ to $\boldsymbol{\mathring{I}}_{i,N_{lom}/N_{lm}}$. 
\begin{figure*}
	\begin{equation}\label{DEQ}
		\begin{aligned}
			\boldsymbol{W}_{RF}&=\underset{\boldsymbol{W}_{LO}}{\underbrace{\left[ \begin{matrix}{}
						e^{j\phi _1}\boldsymbol{\mathring{I}}_{1,1}&		\mathbf{0}&		\cdots&		\mathbf{0}&		\mathbf{0}\\
						\mathbf{0}&		e^{j\phi _1}\boldsymbol{\mathring{I}}_{1,2}&		\cdots&		\mathbf{0}&		\mathbf{0}\\
						\vdots&		\vdots&		\ddots&		\vdots&		\vdots\\
						\mathbf{0}&		\mathbf{0}&		\cdots&		e^{j\phi _{N_{lon}}}\boldsymbol{\mathring{I}}_{N_{lon},N_{lom}/N_{lm}-1}&		\mathbf{0}\\
						\mathbf{0}&		\mathbf{0}&		\cdots&		\mathbf{0}&		e^{j\phi _{N_{lon}}}\boldsymbol{\mathring{I}}_{N_{lon},N_{lom}/N_{lm}}\\
					\end{matrix} \right] }}\times \underset{\boldsymbol{W}_{LC}}{\underbrace{\left[ \begin{matrix}
						\mathbf{1}_{N_{lm}}&		\mathbf{0}&		\cdots&		\mathbf{0}\\
						\mathbf{0}&		\mathbf{1}_{N_{lm}}&		\cdots&		\mathbf{0}\\
						\vdots&		\vdots&		\ddots&		\vdots\\
						\mathbf{0}&		\mathbf{0}&		\cdots&		\mathbf{1}_{N_{lm}}\\
					\end{matrix} \right] }}
			\\
			&=\text{blkdiag}\left\{ e^{j\phi _1}\boldsymbol{\mathring{I}}_{1,1}\mathbf{1}_{N_{lm}},e^{j\phi _1}\boldsymbol{\mathring{I}}_{1,2}\mathbf{1}_{N_{lm}},\cdots ,e^{j\phi _{N_{lon}}}\boldsymbol{\mathring{I}}_{N_{lon},N_{lom}/N_{lm}-1}\mathbf{1}_{N_{lm}},e^{j\phi _{N_{lon}}}\boldsymbol{\mathring{I}}_{N_{lon},N_{lom}/N_{lm}}\mathbf{1}_{N_{lm}} \right\} ,
		\end{aligned}
	\end{equation}
\end{figure*}

Due to this specific structure, Eq. (\ref{CoreProblem}) can be expressed as
\begin{equation}
	\begin{aligned}
		&g\left[ \boldsymbol{W}_{BB},\boldsymbol{W}_{LO} \right] 
		\\
		&\overset{(1)}{=} \left\| \boldsymbol{W}_{opt}-\boldsymbol{\hat{W}}_{LO}\boldsymbol{\mathring{W}}_{LO}\boldsymbol{W}_{LC}\boldsymbol{W}_{BB} \right\| _F
		\\
		&\overset{(2)}{=} \sum_{n=1}^{N_{RF}^{r}} \left\| \left[ \boldsymbol{W}_{opt} \right] _{\left( n-1 \right) N_{lm}+1:nN_{lm},:} \right.
		\\
		&\quad \ \ \ \ \ \ \ \ \ \  \left. -\left[ \boldsymbol{\hat{W}}_{LO}\boldsymbol{\mathring{W}}_{LO}\boldsymbol{W}_{LC}\boldsymbol{W}_{BB} \right] _{\left( n-1 \right) N_{lm}+1:nN_{lm},:} \right\| _F
		\\
		&\overset{(3)}{=} \sum_{n=1}^{N_{RF}^{r}} \left\| \left[ \boldsymbol{W}_{opt} \right] _{\left( n-1 \right) N_{lm}+1:nN_{lm},:} \right.
		\\
		&\quad \ \  \ \ \ \ \ \ \ \ \ \ \ \ \ \ \ \ \ \ \  \left. -e^{j\hat{\phi}_{\mathring{i}}}\boldsymbol{\mathring{I}}_{\mathring{i},\mathring{j}}\mathbf{1}_{N_{lm}}e^{j\left( \phi _{\mathring{i}}-\hat{\phi}_{\mathring{i}} \right)} \left[ \boldsymbol{W}_{BB} \right] _{n,:} \right\| _2
		\\
		&\overset{(4)}{=} \left\| \boldsymbol{W}_{opt}-\boldsymbol{\hat{W}}_{LO}\boldsymbol{W}_{LC}\boldsymbol{\mathring{W}}_{BB}\boldsymbol{W}_{BB} \right\| _F
		\\
		&\overset{}{=} \left\| \boldsymbol{W}_{opt}-\boldsymbol{\hat{W}}_{LO}\boldsymbol{W}_{LC}\boldsymbol{\hat{W}}_{BB} \right\| _F = g\left[ \boldsymbol{\hat{W}}_{BB},\boldsymbol{\hat{W}}_{LO} \right] ,
	\end{aligned}
\end{equation}
where in $\overset{(1)}{=}$, $\boldsymbol{W}_{LO}$ and $\boldsymbol{\hat{W}}_{LO}$ denote the block diagonal matrix constructed from infinite $\phi_i$ and finite $\hat{\phi}_i$ resolution phase shifts, respectively. Since their diagonal blocks are aligned, we can define $\boldsymbol{\mathring{W}}_{LO}=\boldsymbol{\hat{W}}_{LO}^{-1}\boldsymbol{W}_{LO}$, and the phase of each diagonal element satisfies
\begin{equation}
	\begin{aligned}
		\text{arg}\left\{ \left[ \boldsymbol{\mathring{W}}_{LO} \right] _{k,k} \right\}& =\text{arg}\left\{ \left[ \boldsymbol{\hat{W}}_{LO}^{-1} \right] _{k,k}\left[ \boldsymbol{W}_{LO} \right] _{k,k} \right\} 
		\\
		&=\phi _{\lceil \frac{k}{N_{lom}} \rceil}-\hat{\phi}_{\lceil \frac{k}{N_{lom}} \rceil},
	\end{aligned}
\end{equation}
where $k=1,2,\cdots ,N_r$. In $\overset{(2)}{=}$, we transform the Frobenius norm of the original matrix expression into the sum of Frobenius norms of multiple block matrix subproblems. Due to the block-diagonal structure of $\bm{W}_{RF}$, these subproblems are decoupled and thus equivalent to solving each subproblem independently. In $\overset{(3)}{=}$, $\mathring{i}=\lceil n\frac{N_{lm}}{N_{lom}} \rceil $ and $\mathring{j}=\left( \left( n-1 \right) \ \text{mod\ }\frac{N_{lom}}{N_{lm}} \right) +1$. Since both $\bm{W}_{LC}$ and $\bm{\hat{W}}_{LO}$ are diagonal matrices and $N_{lom}$ divides $N_{lm}$, the effect of finite resolution LO phase shifts $e^{j\left( \phi _{\mathring{i}}-\hat{\phi}_{\mathring{i}} \right)}$ can be equivalently represented as multiplying each row of $\bm{W}_{BB}$ by $e^{j\left( \phi _{\mathring{i}}-\hat{\phi}_{\mathring{i}} \right)}$ after basic mathematical reformulation. In $\overset{(4)}{=}$, we reformulate the sum of the Frobenius norm into a consolidated norm expression. 


Furthermore, the results demonstrate that for the optimal $\boldsymbol{W}_{LO}\left(\left\lbrace \phi_i \right\rbrace  \right) $ and $\boldsymbol{W}_{BB}$ corresponding to $\min g\left[\boldsymbol{W}_{LO},\boldsymbol{W}_{BB} \right] $ under infinite-resolution conditions, and for any finite-resolution realization $\boldsymbol{\hat{W}}_{LO}\left(\left\lbrace \hat{\phi_i} \right\rbrace  \right) $, there exists a compensation diagonal matrix $\boldsymbol{\mathring{W}}_{BB}$ with $\text{arg}\left\{ \left[ \boldsymbol{\mathring{W}}_{BB} \right] _{k,k} \right\} =\phi _{\lceil kN_{lm}/N_{lom} \rceil}-\hat{\phi}_{\lceil kN_{lm}/N_{lom} \rceil}$ such that $\boldsymbol{\hat{W}}_{BB}=\boldsymbol{\mathring{W}}_{BB}\bm{W}_{BB}$ and the objective function satisfies $\min g[\boldsymbol{W}_{LO},\boldsymbol{W}_{BB}] = \min g[\boldsymbol{\hat{W}}_{LO},\boldsymbol{\hat{W}}_{BB}]$, which ultimately enables finite-resolution LO to achieve performance equivalent to that of infinite-resolution systems.

\subsubsection{Simplified Algorithm}

Since these $N_{RF}^{r}$ subproblems are decoupled and thus equivalent to solving each subproblem independently due to the block-diagonal structure of $\bm{W}_{RF}$, Eq. (\ref{CoreProblem}) can be expressed as

\begin{equation}
	\begin{aligned}
		&g\left[ \boldsymbol{W}_{BB},\boldsymbol{W}_{LO} \right]= \sum_{n=1}^{N_{RF}^{r}} g_n\left[ \left[ \boldsymbol{W}_{BB}\right] _{n,:} ,{\phi}_{\mathring{i}}\right]
		\\
		&\overset{}{=} \sum_{n=1}^{N_{RF}^{r}} \left\| \left[ \boldsymbol{W}_{opt} \right] _{\left( n-1 \right) N_{lm}+1:nN_{lm},:} \right.
		\\
		&\quad \ \  \ \ \ \ \ \ \ \ \ \ \ \ \ \ \ \ \ \ \  \  \ \ \ \ \ \left. -e^{j{\phi}_{\mathring{i}}}\boldsymbol{\mathring{I}}_{\mathring{i},\mathring{j}}\mathbf{1}_{N_{lm}} \left[ \boldsymbol{W}_{BB} \right] _{n,:} \right\| _F.
	\end{aligned}
\end{equation}

For each subproblem $\min g_n\left[ \left[ \boldsymbol{W}_{BB}\right] _{n,:},{\phi}_{\mathring{i}} \right]$, we have
\begin{equation}
	\begin{aligned}
		\left[ \boldsymbol{W}_{BB} \right] _{n,:}&=\left( e^{j\phi _{\mathring{i}}}\boldsymbol{\mathring{I}}_{\mathring{i},\mathring{j}}\mathbf{1}_{N_{lm}} \right) ^{\dag}\left[ \boldsymbol{W}_{opt} \right] _{\left( n-1 \right) N_{lm}+1:nN_{lm},:}
		\\
&=\frac{1}{N_{lm}}e^{-j\phi _{\mathring{i}}}\mathbf{1}_{N_{lm}}^{T}\boldsymbol{\mathring{I}}_{\mathring{i},\mathring{j}}^{H}\left[ \boldsymbol{W}_{opt} \right] _{\left( n-1 \right) N_{lm}+1:nN_{lm},:}.
	\end{aligned}
\end{equation}

The above result allows the MP pseudoinverse of the original $N_r \times N_{RF}^r$ matrix in Eq. (\ref{step1}) be decomposed into the product of $N_{RF}^r$ smaller matrices, entirely without iterative computation. This approach remains valid for discrete phase shifts $\hat{\phi} _{\mathring{i}}$.

\begin{figure}
	\centering
	\includegraphics[width=0.42\textwidth]{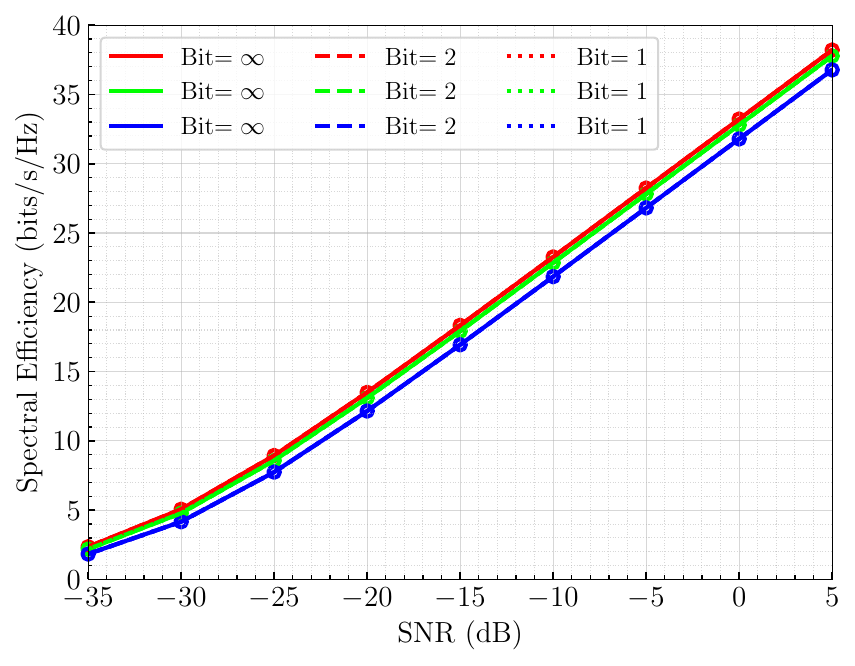}
	\caption{Spectral efficiency under $N_s=3$ and $N_{lom}=6$. Red line: $N_{lm}=1$. Green line: $N_{lm}=3$. Blue line: $N_{lm}=6$.}
	\label{PYLC_DS_Ns3_Nlom6_Nlm136}
\end{figure}

\begin{figure}
	\centering
	\includegraphics[width=0.42\textwidth]{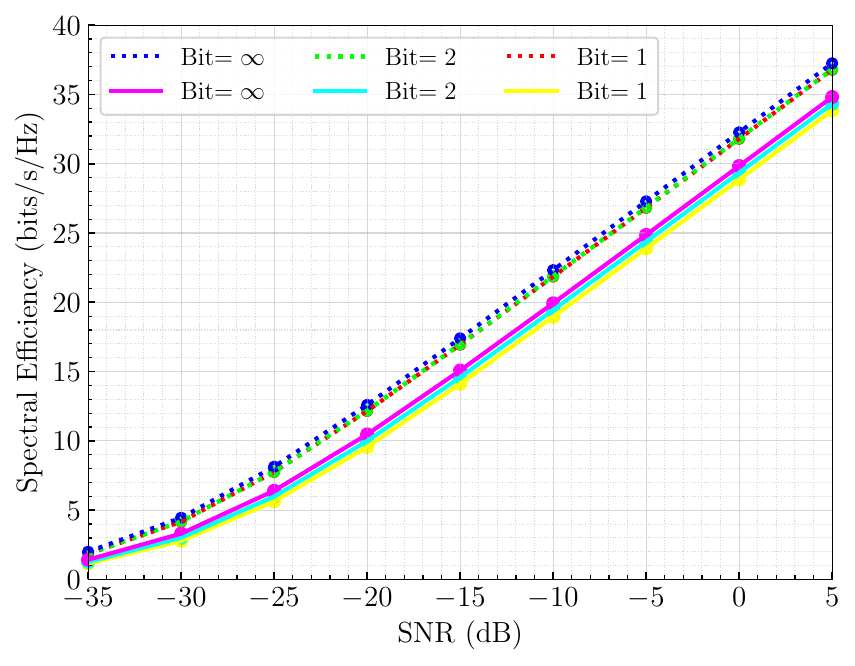}
	\caption{Spectral efficiency under $N_s=3$ and $N_{lom}=6$. Dotted line : $N_{lm}=4$. Solid line : $N_{lm}=12$.}
	\label{PYLC_DS_Ns3_Nlom6_Nlm412_B}
\end{figure}

\section{Simulation Results}\label{T5} 

In this section, we numerically evaluate the performance of the proposed algorithms. Data streams are transmitted from a transmitter equipped with $N_t = 144$ antennas to a receiver with $N_{lon} = 36$ Cell $\&$ LO blocks. The transmitter and the LO-D architecture are equivalently modeled as employing uniform square planar arrays, while the antenna array response vector for the LO-S architecture is derived based on the analysis presented in Sec. \ref{T31}. The channel parameters are configured with $N_{cl} = 5$ clusters and $N_{ray} = 10$ rays per cluster, where each cluster has an average power of $\sigma^2_{\alpha,i} = 1$. We focus exclusively on the impact of receiver-side reuse architectures. Therefore, the transmitter is assumed to employ ideal precoding, satisfying $\bm{F}_{opt} = \bm{F}_{RF} \bm{F}_{BB}$. The azimuth and elevation angles of departure and arrival follow the Laplacian distribution, with mean angles uniformly distributed over $[0, 2\pi)$ and an angular spread of 10 degrees. The antenna elements are modeled as point antennas, and all simulation results are averaged over 2000 channel realizations. For all alternating minimization algorithms proposed, the initial phases of the analog precoder $\bm{W}_{LO}$ are drawn from a uniform distribution over $[0, 2\pi)$. We assume that all $N_{lon}$ Cell $\&$ LO blocks are spaced at a constant interval $d = \lambda/2$, and the intra-block antenna spacing is fixed at $\hat{d} = 0.1d$. Consequently, the phases $\varphi _{i,k}$ of the diagonal elements in $\boldsymbol{\tilde{I}}_i$ is given by
\begin{equation}
	\begin{aligned}
\varphi _{i,k}=0.1\pi \left( k-\frac{N_{lom}-1}{2} \right) ,\ k=0,1,\cdots ,N_{lom}-1.
	\end{aligned}
\end{equation}

\subsection{Performance Under LO Reuse}

Figure \ref{LO_DS_Ns3} illustrates the spectral efficiency performance for $N_s = 3$. The adoption of LO reuse increases the number of antenna elements, with $N_{lom}$ rising from 1 to 6 across $N_{lon}$ cell $\&$ LO blocks, and the total receiver antenna count $N_r$ increasing from 36 (without LO reuse) to 216, leading to a corresponding improvement in spectral efficiency. Notably, this performance improvement is preserved even under finite LO resolution.

\subsection{Performance Under LC Reuse}
\subsubsection{Performance Under Finite Resolution}

Figures \ref{LCReuse} present the spectral efficiency performance under different LO reuse depths $N_{lom}$ and APD reuse depths $N_{lm}$. The cyan dashed line represents the configuration with $N_{lm} = 1$ (no APD reuse) and $N_{lom} = 6$ (deep LO reuse), which corresponds to an ideal combiner and achieves the highest performance. The magenta dotted line corresponds to $N_{lm} = 1$ and $N_{lom} = 1$ (no LO reuse and no LC reuse), also equivalent to an ideal combiner, representing the current work \cite{gong2025rydberg,zhu2025raq,atapattu2025detection,song2025csi,cui2025mimo,xu2025channel,cui2024multi,yuan2025electromagnetic,xu2024design} performance for Rydberg MIMO receivers with $N_r=N_{lon} = 36$, i.e., the optimal performance achievable without LO reuse. For a fixed LO reuse depth $N_{lom}$, the spectral efficiency decreases as the APD reuse depth $N_{lm}$ increases. The blue curves in Fig. \ref{LCReuse} reveals that for a fixed number of laser chains ($N_{RF}^r = 6$ for Fig. \ref{LCReuse}(a) and Fig. \ref{LCReuse}(b)), higher LO reuse depth $N_{lom}$ leads to improved performance after APD reuse. This indicates that without increasing the number of APD, system performance can be enhanced by deploying deeper LO reuse $N_{lom}$. Furthermore, comparing the solid and magenta dotted lines across the two subfigures shows that employing both LO and APD reuse achieves better performance with fewer APD than the LO-D architecture, i.e., the conventional antenna design approach in current work. The above result validates the effectiveness of the proposed reuse design and precoding algorithm.

\subsubsection{Performance Under Infinite Resolution}

In Fig. \ref{PYLC_DS_Ns3_Nlom6_Nlm136}, we present the spectral efficiency results for the special scenario introduced in Sec. \ref{T34} where the APD reuse depth is divisible by the LO reuse depth. For an LO reuse depth of 6, the red, green, and blue curves correspond to APD reuse depths of 1, 3, and 6, respectively. Under these configurations, the performance under low-bit quantization matches that of the continuous-phase scenario.

In Fig. \ref{PYLC_DS_Ns3_Nlom6_Nlm412_B}, we present results for the more general case where the APD reuse depth is not divisible by the LO reuse depth. For an LO reuse depth of 6, the solid and dashed lines represent APD reuse depths of 12 and 4, respectively. In such scenarios, higher-resolution quantization yields superior performance.

\begin{figure}
	\centering
	\includegraphics[width=0.42\textwidth]{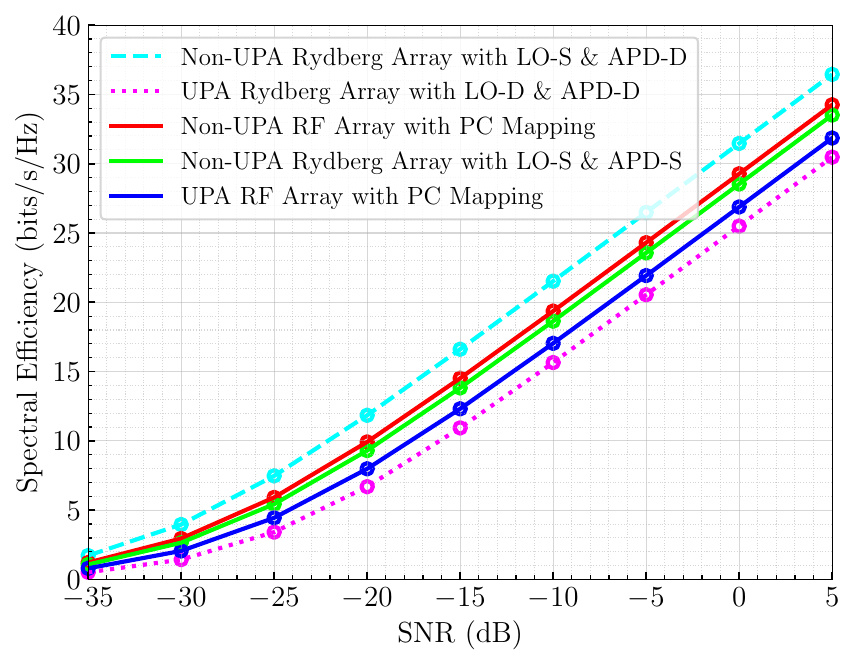}
	\caption{Spectral efficiency under $N_s=3$, $N_{lom}=4$ and $N_{RF}^r=12$.}
	\label{PYCompare1_Ns3Nt144N_lon36N_lom4H_Num3000}
\end{figure}

\begin{figure}
	\centering
	\includegraphics[width=0.42\textwidth]{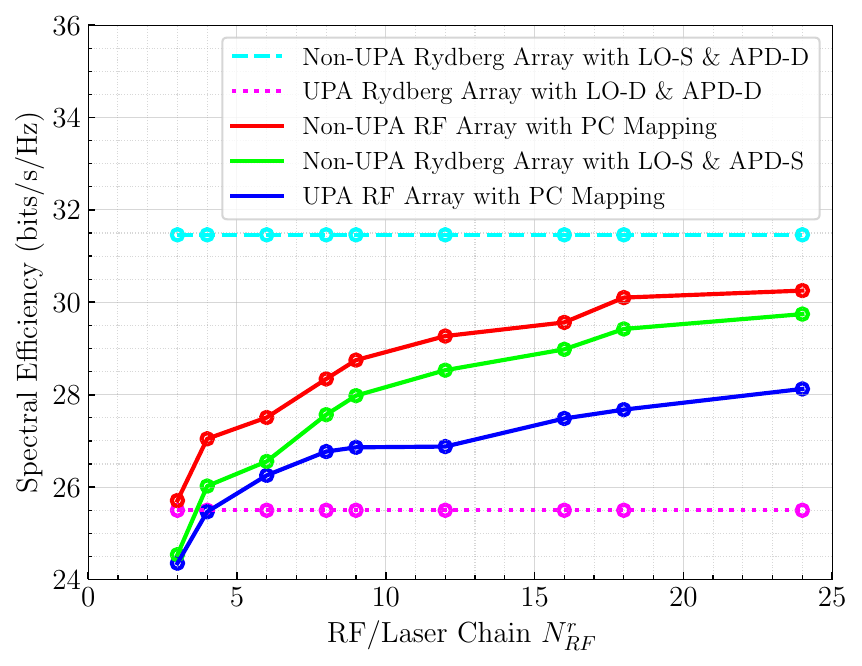}
	\caption{Spectral efficiency under $N_s=3$, $N_{lom}=4$ and $\text{SNR}=0$.}
	\label{PYCompare2_Ns3Nt144N_lon36N_lom4H_Num3000}
\end{figure}

\subsection{Performance Comparison: Rydberg Reuse Arrays vs Traditional RF Antennas with PC Mapping}

Figure \ref{PYCompare1_Ns3Nt144N_lon36N_lom4H_Num3000} and Figure \ref{PYCompare2_Ns3Nt144N_lon36N_lom4H_Num3000} focus on comparing the reuse performance between Rydberg arrays and traditional RF antennas within a PC mapping framework. Since the Rydberg multiplexed array is inherently Non-UPA, we include for comparison both a UPA with equal antenna count ($N_r = N_{lon} \times N_{lom} = 36 \times 4$) and a non-UPA with identical layout to the Rydberg array in the conventional antenna evaluation. The transmitter is assumed to employ ideal precoding in all cases, enabling a focused comparison of receiver-side performance. Although individual Rydberg sensors exhibit higher sensitivity, we assume equal signal-to-noise ratios for both systems in this analysis to isolate the advantages arising solely from the inherent reuse mapping strategy of Rydberg arrays.

Figure \ref{PYCompare1_Ns3Nt144N_lon36N_lom4H_Num3000} illustrates the variation of array performance with SNR, while Figure \ref{PYCompare2_Ns3Nt144N_lon36N_lom4H_Num3000} shows the spectral efficiency as a function of the number of RF chains (or laser chains for Rydberg systems). The red and blue curves in both figures represent, respectively, a standard UPA with spacing $d$ and a Non-UPA layout matching the Rydberg receiver's geometry. As reference benchmarks, the cyan dashed line represents the configuration with $N_{lm} = 1$ (no APD reuse), $N_{lom} = 4$ and $N_r=36\times 4$, corresponding to an ideal combiner and achieving the highest performance. The magenta dotted line corresponds to $N_{lm} = 1$ and $N_{lom} = 1$ (no LO and LC reuse), also equivalent to an ideal combiner, representing the current state-of-the-art performance for Rydberg MIMO receivers with $N_r = N_{lon} = 36$, the optimal performance achievable without LO reuse. The results show that, under equal antenna count $N_r$ and with either equal RF/laser chain numbers $N_{RF}^r$ or equal SNR conditions, the Rydberg reuse array outperforms the traditional UPA but is surpassed by the non-UPA with PC mapping configuration. This performance hierarchy stems directly from the inherent reuse strategy of the Rydberg array. Compared with the PC mapping in conventional antennas, the proposed reuse architecture employs fewer phase shifters that are shared among antenna elements.

\section{Conclusion}\label{T6}
This paper systematically establishes a theoretical and optimization framework for multiplexed array design in Rydberg atom sensor-based hybrid massive MIMO systems. We introduce the two reuse strategy encompassing both LO and APD sharing configurations. A hybrid precoding algorithm employing alternating optimization is developed, with numerical simulations confirming significant performance advantages of the proposed architectures over conventional arrays.

\appendices

\section{}\label{A1}
\subsection{The Derivation for Infinite Resolution $\phi_{i}$}\label{A11}
Denote $\bm{Y}=[\boldsymbol{W}_{opt}]_{(i-1)N_{lom}+1:iN_{lom},:}$ and $\bm{X}=\boldsymbol{\tilde{I}}_i [\boldsymbol{W}_{LC} \boldsymbol{W}_{BB}]_{(i-1)N_{lom}+1:iN_{lom},:}$, the problem in Eq. (\ref{IWLO}) is given by
\begin{equation}
	\begin{aligned}
&\lVert \boldsymbol{Y}-e^{j\phi _i}\boldsymbol{X} \rVert _{F}^{2}
\\
=&\text{Tr}\left[ \left( \boldsymbol{Y}-e^{j\phi _i}\boldsymbol{X} \right) ^H\left( \boldsymbol{Y}-e^{j\phi _i}\boldsymbol{X} \right) \right] 
\\
=&\lVert \boldsymbol{Y} \rVert _{F}^{2}+\lVert \boldsymbol{X} \rVert _{F}^{2}-e^{-j\phi _i}\text{Tr}\left[ \boldsymbol{X}^H\boldsymbol{Y} \right] -e^{j\phi _i}\text{Tr}\left[ \boldsymbol{Y}^H\boldsymbol{X} \right] 
\\
=&\lVert \boldsymbol{Y} \rVert _{F}^{2}+\lVert \boldsymbol{X} \rVert _{F}^{2}-2\left| \text{Tr}\left[ \boldsymbol{X}^H\boldsymbol{Y} \right] \right|\cos \left( \phi _i-\theta _i \right) ,
	\end{aligned}
\end{equation}
where $\theta _i=\arg \left\{ \text{Tr}\left[ \boldsymbol{X}^H\boldsymbol{Y} \right] \right\} $. Thus, for $\phi_{i} \in \left[0,2\pi \right] $, we have

\begin{equation}
	\begin{aligned}
		\phi_i &= \arg\min_{\phi_i} \lVert \boldsymbol{Y} - e^{j\phi_i}\boldsymbol{X} \rVert_F^2 \\
		&= \arg\max_{\phi_i} \ \cos(\phi_i - \theta_i) \\
		&= \theta_i = \arg \left\{ \text{Tr}\left[ \boldsymbol{X}^H\boldsymbol{Y} \right] \right\},
	\end{aligned}
\end{equation}
where substituting the specific matrices $\boldsymbol{X}$ and $\boldsymbol{Y}$ yields Eq. (\ref{IWLO}).

\subsection{The Derivation for Finite Resolution $\hat{\phi}_{i}$}\label{A12}

For $B$ bits of resolution $\hat{\phi}_{i} \in \mathcal{B}$, we have 
\begin{equation}
	\begin{aligned}
		\hat{\phi}_i &= \arg\min_{\hat{\phi}_{i} \in \mathcal{B}} \lVert \boldsymbol{Y} - e^{j\hat{\phi}_i}\boldsymbol{X} \rVert_F^2 \\
		&= \arg\max_{\hat{\phi}_{i} \in \mathcal{B}} \ \cos(\hat{\phi}_i - \theta_i) \\
		&= \arg\min_{\hat{\phi}_{i} \in \mathcal{B}} \ \left| \hat{\phi} - \theta_i\right| ,
	\end{aligned}
\end{equation}
where the properties of the cosine function ensure that the phase value $\hat{\phi}_i \in \mathcal{B}$ closest to $\theta_i$ constitutes the optimal solution under $B$-bit resolution. Moreover, since the problem formulated in Eq. (\ref{SubProblem2_1}) decomposes into multiple independent subproblems, each associated with a distinct $\hat{\phi}_i$, the set of $\hat{\phi}_i$ values obtained in each iteration via this nearest-neighbor rule collectively produces the optimal $\bm{W}_{LO}^{[p]}$ under the given resolution constraint.

\ifCLASSOPTIONcaptionsoff
\newpage
\fi

\normalem
\bibliographystyle{IEEEtran}
\bibliography{myref}





%

%
%
%




\end{document}